\documentclass[%
reprint,
superscriptaddress,
 amsmath,amssymb,
 aps,
]{revtex4-2}

\usepackage[caption=false]{subfig}
\usepackage{array}[=2016-10-06]
\usepackage{tikz}
\usepackage{booktabs} 
\usepackage{graphicx}
\usepackage{dcolumn}
\usepackage{siunitx}
\usepackage{subcaption}
\usepackage{bm}
\usepackage{hyperref}
\usepackage{verbatim}
\usepackage{cleveref}
\usepackage{threeparttable}

\begin{document}

\preprint{APS/123-QED}

\title{Mass-Dependent Non-Extensivity in Tsallis Blast-Wave Fits to Identified Hadron $p_T$ Spectra at RHIC and LHC}

\author{C. Y. Tsang}
 \email{ctsang@anl.gov}
\affiliation{Argonne National Laboratory, Argonne, Illinois 60439}

\author{Z. Xu}
\affiliation{Kent State University, Kent, Ohio 44242}

\date{\today}

\begin{abstract}

We analyze identified-hadron transverse-momentum spectra from STAR Au+Au and ALICE Pb+Pb collisions over $\sqrt{s_{NN}} = 7.7$~GeV--$5.02$~TeV using an extended Tsallis Blast-Wave (TBW) framework, which includes a non-extensivity parameter $q$ to quantify the degree of incomplete thermal equilibrium. Conventionally, either a common value of $q$, or two separate $q$ values, one for mesons and one for baryons, are used to describe particle spectra in the TBW framework, with the latter being referred to as TBW4 in Ref.~\cite{Chen:2020zuw}. This work extends the TBW framework by studying the dependence of $q$ on different kinds of particles in detail. Fits allowing independent non-extensivity parameters $q$ for each species reveal a systematic correlation between $q$ and particle mass, except for quarkonia. Motivated by this trend, we introduce two new parameterizations: TBW5, which posits that $q$ depends linearly on particle mass, and TBW6, which allows the $q$ intercepts for mesons and baryons to differ. Across all energies and centralities considered in this study, TBW5 improves $\chi^{2}/\mathrm{NDF}$ relative to the TBW4 fit in 74\% of the datasets, while TBW6 shows improvement in 93\% of the datasets. They perform especially well in central collisions. These results demonstrate a robust mass ordering in non-equilibrium behavior at kinetic freeze-out and provide a more accurate description of hadron spectra from RHIC to LHC energies.

\end{abstract}

\maketitle


\section{\label{sec:intro}Introduction}

Relativistic heavy-ion collisions generate an extremely hot, dense fireball that exists for a brief moment as the deconfined quark-gluon plasma (QGP)~\cite{Gyulassy:2004zy,Heinz:2013th,Busza:2018rrf,Shuryak:2017rcp}. As the fireball expands, it cools and eventually reaches the chemical freeze-out phase, where inelastic collisions stop and particle yields are fixed. A later and cooler kinetic freeze-out phase freezes elastic interactions, fixing the momentum distributions of the emitted particles~\cite{STAR:2005gfr}. The transverse-momentum ($p_T$) spectra measured at this stage therefore encode valuable information about the system's properties at final freeze-out.

The Blast-Wave (BW) model has long been used to describe these $p_T$ spectra. It assumes that the fireball (i) is in thermal equilibrium and (ii) follows a simple collective radial flow profile as its local velocity distribution~\cite{Becattini:1997rv, schnedermann1994hydrodynamical}. This picture works well at low $p_T$~\cite{STAR:2008med, STAR:2017sal}, but it falls short at higher momenta. To bridge that gap, Tsallis statistics were introduced~\cite{Tang:2008ud, Shao:2009mu, Tang:2011xq, Ristea:2013ara, Chen:2020zuw,Ajaz:2024uvd,Lao:2017skd}. By adding a non‑extensivity parameter $q$, the Tsallis distribution interpolates smoothly between a Boltzmann-Gibbs energy distribution (recovered when $q=1$) and a power‑law tail characteristic of slight departures from equilibrium ($q>1$), thereby providing a unified description of both soft and hard particle production~\cite{Tang:2008ud}.

When the Tsallis function is applied systematically to identified-hadron spectra from ALICE (Pb+Pb at $\sqrt{s_{NN}}=2.76$ and \SI{5.02}{TeV}) and STAR (Au+Au from BES-I and other runs), a single $q$ value for all species does not give an optimal fit~\cite{Chen:2020zuw}. Ref.~\cite{Chen:2020zuw} suggests that mesons and baryons may experience different degrees of thermalization. Therefore, two separate non-extensivity parameters are introduced: $q_{\rm meson}$ for mesons and $q_{\rm baryon}$ for baryons, while keeping temperature, radial flow profile, and the other freeze-out parameters common to every species. The two-$q$ scheme markedly improves the description of both the STAR Au+Au and the ALICE Pb+Pb spectra~\cite{Chen:2020zuw}.

Even though separating the $q$ values for mesons and baryons helps, it is not clear whether this way of ascribing $q$ values to species is optimal. In the present work, we extend this train of thought by exploring alternative ways to let the Tsallis non-extensivity parameter vary among identified hadrons. Rather than a single $q$ for all mesons and a single $q$ for all baryons, we test parameterizations where $q$ depends on intrinsic particle characteristics, while keeping the total number of fit degrees of freedom unchanged or allowing only a modest increase. By weighing improvements in fit quality against the penalty of extra parameters, we assess whether a more nuanced $q$-distribution is justified and what it can reveal about non-equilibrium dynamics at kinetic freeze-out.

\section{Background}

The BW model describes particle spectra as a result of emission from a locally moving, perfectly thermalized fireball~\cite{Becattini:1997rv, schnedermann1994hydrodynamical}. The initial radial velocity profile of the fireball is assumed to be

\begin{equation}
\beta(r) = \beta_{S}\left(\frac{r}{R}\right)^{n},
\label{eq:velocityProfile}
\end{equation}

where $\beta_{S}$ and $n$ are fit parameters. $R$ is the radius of the fireball, but this value can be absorbed into the normalization constant of the BW function via a change of variable, so its exact value does not matter. Throughout this analysis, $n$ is fixed to unity. Under the additional assumption of thermal equilibrium, particle energy in the fireball's local rest frame follows a Boltzmann-Gibbs energy distribution. After transforming back to the laboratory frame, the resulting invariant yield reads

\begin{equation}
\begin{split}
\frac{1}{2\pi p_T}\frac{d^2N}{dydp_T}
\propto \int_0^R r\,m_T\,I_0\!\left(\frac{p_T\sinh\rho(r)}{T_{\text{kin}}}\right) \\
\times K_1\!\left(\frac{m_T\cosh\rho(r)}{T_{\text{kin}}}\right)\,dr,
\end{split}
\label{eq:BW}
\end{equation}

with $I_{0}$ and $K_{1}$ being the modified Bessel functions of the first and second kind, respectively; $m_T=\sqrt{p_T^{2}+m_0^{2}}$ with $m_0$ being the rest mass of the particle; and $\rho(r)=\tanh^{-1}\!\bigl[\beta(r)\bigr]$. Since BW with a common kinetic temperature $T_{\rm kin}$ and velocity profile can simultaneously describe the $p_T$ spectra of pions, kaons, and protons at relatively low $p_T$ ($\leq\sim\SI{1.3}{GeV/c}$)~\cite{STAR:2008med, STAR:2017sal}, it suggests that the fireball has reached, or is at least close to, thermal equilibrium before hadronization. The flow profile is often quantified by the average flow velocity $\langle\beta\rangle = 2\,\beta_{S}/(2+n)$~\cite{Ristea:2013ara}. 

However, the BW description breaks down at higher $p_T$. It fails to account for the power-law tails observed experimentally, indicating that non-equilibrium effects become important. In fact, it is known that high $p_T$ spectra receive contributions from non-thermal sources, such as jets or hadrons from hard scattering~\cite{STAR:2004bgh, Wilk:1999dr, Wong:2015mba, Urmossy:2011xk}, which is consistent with the observed power-law tails. 

Consequently, a more general statistical framework is required to describe the $p_T$ spectra over a wide $p_T$ range. Tsallis statistics extends the Boltzmann distribution by introducing a non-extensivity parameter $q$ that quantifies deviations from equilibrium~\cite{De:2007zza, Wilk:2009nn, Alberico:1999nh, Osada:2008sw}. The Tsallis distribution for the particle spectrum in the source rest frame reads  

\begin{equation}
f(m_T)\propto
\left[1+\frac{q-1}{T}\,m_T\right]^{-\frac{1}{\,q-1}} ,
\label{eq:Tsallis}
\end{equation}

where $q=1$ reduces the expression to the ordinary Boltzmann distribution and larger values $q>1$ correspond to increasingly non-equilibrated behavior. Replacing the Boltzmann factor in the BW model with Eq.~\eqref{eq:Tsallis} yields the Tsallis Blast-Wave (TBW) model, whose invariant yield can be written as  

\begin{widetext}
\begin{equation} 
\begin{split} 
  \frac{1}{2\pi p_T}\frac{d^2N}{dydp_T} \propto \int_{-y_b}^{y_b}\int_{-\pi}^{\pi}\int_{0}^{R} &m_Tre^{\sqrt{y_b - y}}\cosh{y}\big(1+\frac{q-1}{T}(m_T\cosh{y}\cosh{\rho} \\ 
  &-p_T\sinh{\rho}\cos{\phi})\big)^{-\frac{1}{q-1}}drd\phi dy. 
\end{split} 
\end{equation} 
\end{widetext}

Here $y_b$ denotes the beam rapidity. The TBW model, introduced in Refs.~\cite{Tang:2008ud, Shao:2009mu, Tang:2011xq, Ristea:2013ara, Chen:2020zuw}, successfully describes a myriad of identified-hadron spectra simultaneously over a relatively wide $p_T$ range, capturing both the low-$p_T$ region governed by collective flow and the intermediate-$p_T$ tail reflecting nonequilibrium dynamics. Thus, while the conventional BW approach is a useful tool under equilibrium conditions, the Tsallis-based extension provides a more complete framework for interpreting $p_T$ spectra in relativistic heavy-ion collisions.

Ref.~\cite{Chen:2020zuw} found that the description of particle spectra can be improved if separate $q$ values are used for mesons and baryons, rather than a single shared $q$ value for all particles. However, other ways of assigning $q$ values to hadrons have not been explored. In the following sections, we study the optimal assignment of $q$ values to different particle species.

\begin{table*}
    \centering
    \begin{threeparttable}
        \caption{Particle, centrality and data reference list for all beam energies considered in this manuscript. A tick indicates that data exist for the corresponding centrality bin, while a cross indicates that the particle is not included for that bin. When a percentage range is shown instead of a tick or cross, data from the stated centrality range are used as a substitute for that centrality bin. This approximation is required for some species due to limited data availability.}
    \label{tab:centDist}
    \label{tab:particleList}
    \begin{tabular}{cl|c|c|c|c|c|c|c|c|c|c}
    \toprule
        $\sqrt{s_{NN}}$ (GeV) & \multicolumn{1}{c}{Particles} & \multicolumn{9}{c}{Centralities} & Ref. \\
        \hline
        7.7, 11.5 & \multicolumn{1}{c}{} &  \multicolumn{1}{c}{0-5\%} & 
        \multicolumn{1}{c}{5-10\%} & \multicolumn{1}{c}{10-20\%} & \multicolumn{1}{c}{20-30\%} & \multicolumn{1}{c}{30-40\%} & \multicolumn{1}{c}{40-60\%} & \multicolumn{1}{c}{60-80\%} & \multicolumn{1}{c}{} & \multicolumn{1}{c}{} &  \\
        \cline{3-9}
        & $\pi^{\pm},\, K^{\pm},\, p,\, \bar{p}$ & \checkmark & \checkmark & \checkmark & \checkmark & \checkmark & \checkmark & \checkmark & \multicolumn{1}{c}{} & \multicolumn{1}{c}{} & \cite{STAR:2017sal} \\
        \cline{3-9}
        & $\Lambda$, $\bar{\Lambda}$, $\Xi^{\pm}$ & \checkmark & \checkmark & \checkmark & \checkmark & \checkmark & \checkmark & \checkmark & \multicolumn{1}{c}{} & \multicolumn{1}{c}{} & \cite{STAR:2019bjj} \\
        \cline{3-9}
        & $\phi$ & 0-10\% & $\times$ & \checkmark & \checkmark & \checkmark & \checkmark & \checkmark & \multicolumn{1}{c}{} & \multicolumn{1}{c}{} & \cite{STAR:2015vvs} \\
        \cline{1-9}
        19.6, 27, 39 & \multicolumn{1}{c}{} &  \multicolumn{1}{c}{0-5\%} & 
        \multicolumn{1}{c}{5-10\%} & \multicolumn{1}{c}{10-20\%} & \multicolumn{1}{c}{20-30\%} & \multicolumn{1}{c}{30-40\%} & \multicolumn{1}{c}{40-60\%} & \multicolumn{1}{c}{60-80\%} & \multicolumn{1}{c}{} & \multicolumn{1}{c}{} \\
        \cline{3-9}
        &  $\pi^{\pm},\, K^{\pm},\, p,\, \bar{p}$ & \checkmark & \checkmark & \checkmark & \checkmark & \checkmark & \checkmark & \checkmark & \multicolumn{1}{c}{} & \multicolumn{1}{c}{} & \cite{STAR:2017sal}\\
        \cline{3-9}
        &  $\Lambda$, $\bar{\Lambda}$, $\Xi^{\pm}$ & \checkmark & \checkmark & \checkmark & \checkmark & \checkmark & \checkmark & \checkmark & \multicolumn{1}{c}{} & \multicolumn{1}{c}{} & \cite{STAR:2019bjj}\\
        \cline{3-9}
        &  $\Omega^{\pm}$ & 0-10\%  & $\times$ & \checkmark & 20-40\% & $\times$ & \checkmark & $\times$ & \multicolumn{1}{c}{} & \multicolumn{1}{c}{} & \cite{STAR:2015vvs}\\
        \cline{3-9}
        & $\phi$ & 0-10\% & $\times$ & \checkmark & \checkmark & \checkmark & \checkmark & \checkmark & \multicolumn{1}{c}{} & \multicolumn{1}{c}{} & \cite{STAR:2015vvs}\\
        \cline{1-9}
        62.4 & \multicolumn{1}{c}{} & \multicolumn{1}{c}{0-20\%} & \multicolumn{1}{c}{20-40\%} & \multicolumn{1}{c}{40-80\%} & \multicolumn{1}{c}{} & \multicolumn{1}{c}{} & \multicolumn{1}{c}{} & \multicolumn{1}{c}{} \\
        \cline{3-5}
         &  $\pi^{\pm},\, K^{\pm},\, p,\, \bar{p}$ & \checkmark & \checkmark & \checkmark & \multicolumn{1}{c}{} & \multicolumn{1}{c}{} & \multicolumn{1}{c}{} & \multicolumn{1}{c}{} & \multicolumn{1}{c}{} & \multicolumn{1}{c}{} & \cite{STAR:2008med} \\
         \cline{3-5}
         & $K_{S}^{0}$, $\Lambda$\tnote{a}, $\bar{\Lambda}$\tnote{a}, $\Xi^{\pm}$\tnote{a} & \checkmark & \checkmark & \checkmark & \multicolumn{1}{c}{} & \multicolumn{1}{c}{} & \multicolumn{1}{c}{} & \multicolumn{1}{c}{} & \multicolumn{1}{c}{} & \multicolumn{1}{c}{} & \cite{STAR:2010yyv} \\
         \cline{3-5}
         &  $\phi$\tnote{a} & \checkmark & \checkmark & \checkmark & \multicolumn{1}{c}{} & \multicolumn{1}{c}{} & \multicolumn{1}{c}{} & \multicolumn{1}{c}{} & \multicolumn{1}{c}{} & \multicolumn{1}{c}{} & \cite{STAR:2008bgi} \\
        \cline{1-5}
        200 & \multicolumn{1}{c}{} & \multicolumn{1}{c}{0-10\%} & \multicolumn{1}{c}{10-20\%} & \multicolumn{1}{c}{20-40\%} & \multicolumn{1}{c}{40-60\%} & \multicolumn{1}{c}{60-80\%} & \multicolumn{1}{c}{} & \multicolumn{1}{c}{} \\
        \cline{3-7}
         &  $\pi^{\pm},\, p,\, \bar{p}$ & \checkmark & \checkmark & \checkmark & \checkmark & \checkmark & \multicolumn{1}{c}{} & \multicolumn{1}{c}{}& \multicolumn{1}{c}{} & \multicolumn{1}{c}{} & \cite{STAR:2008med}\\
         \cline{3-7}
         &  $K^{\pm}$ & \checkmark & \checkmark & \checkmark & \checkmark & \checkmark & \multicolumn{1}{c}{} & \multicolumn{1}{c}{}& \multicolumn{1}{c}{} & \multicolumn{1}{c}{} & \cite{PHENIX:2013kod}\\
         \cline{3-7}
         & $\Lambda$, $\bar{\Lambda}$, $\Xi^{\pm}$ & 0-5\% & \checkmark & \checkmark & \checkmark & \checkmark & \multicolumn{1}{c}{}& \multicolumn{1}{c}{}& \multicolumn{1}{c}{}& \multicolumn{1}{c}{} & \cite{STAR:2006egk}\\
         \cline{3-7}
         &  $\Omega$+$\bar{\Omega}$ & 0-5\% & $\times$ & \checkmark & \checkmark & $\times$ & \multicolumn{1}{c}{}& \multicolumn{1}{c}{}& \multicolumn{1}{c}{}& \multicolumn{1}{c}{}  & \cite{STAR:2006egk}\\
         \cline{3-7}
         &  $\phi$\tnote{a} & \checkmark & \checkmark & \checkmark & \checkmark & \checkmark & \multicolumn{1}{c}{} & \multicolumn{1}{c}{}& \multicolumn{1}{c}{} & \multicolumn{1}{c}{} & \cite{STAR:2007mum}\\
        \cline{1-7}
         2760 & \multicolumn{1}{c}{} & \multicolumn{1}{c}{0-10\%} & \multicolumn{1}{c}{10-20\%} & \multicolumn{1}{c}{20-40\%} & \multicolumn{1}{c}{40-60\%} & \multicolumn{1}{c}{60-80\%} & \multicolumn{1}{c}{} & \multicolumn{1}{c}{}\\
         \cline{3-7}
         & $\pi^{\pm},\, K^{\pm},\, p,\, \bar{p}$ & \checkmark & \checkmark & \checkmark & \checkmark & \checkmark & \multicolumn{1}{c}{}& \multicolumn{1}{c}{}& \multicolumn{1}{c}{}& \multicolumn{1}{c}{} & \cite{ALICE:2013mez} \\
         \cline{3-7}
         & $\Lambda$ & \checkmark & \checkmark & \checkmark & \checkmark & \checkmark & \multicolumn{1}{c}{}& \multicolumn{1}{c}{}& \multicolumn{1}{c}{}& \multicolumn{1}{c}{} & \cite{ALICE:2013cdo} \\
         \cline{3-7}
         & $\Xi^{\pm}$, $\Omega^{\pm}$ & \checkmark & \checkmark & \checkmark & \checkmark & \checkmark & \multicolumn{1}{c}{}& \multicolumn{1}{c}{}& \multicolumn{1}{c}{}& \multicolumn{1}{c}{} & \cite{ALICE:2013xmt} \\
         \cline{3-7}
         & $\phi$ & \checkmark & \checkmark & \checkmark & \checkmark & \checkmark & \multicolumn{1}{c}{}& \multicolumn{1}{c}{}& \multicolumn{1}{c}{}& \multicolumn{1}{c}{} & \cite{ALICE:2014jbq} \\
         \cline{1-7}
         5020 & \multicolumn{1}{c}{} & \multicolumn{1}{c}{0-5\%} & 
        \multicolumn{1}{c}{5-10\%} & \multicolumn{1}{c}{10-20\%} & \multicolumn{1}{c}{20-30\%} & \multicolumn{1}{c}{30-40\%} & \multicolumn{1}{c}{40-50\%} & \multicolumn{1}{c}{50-60\%} & \multicolumn{1}{c}{60-70\%} & \multicolumn{1}{c}{70-80\%} \\
        \cline{3-11}
         & $\pi^{+}$+$\pi^{-}$, $K^{+}$+$K^{-}$, $p$+$\bar{p}$ & \checkmark & \checkmark & \checkmark & \checkmark & \checkmark & \checkmark & \checkmark & \checkmark & \checkmark & \cite{ALICE:2019hno} \\
         \cline{3-11}
         & $\Lambda+\bar{\Lambda}$, $\Xi+\bar{\Xi}$, $\Omega+\bar{\Omega}$ & 0-10\% & $\times$ & \checkmark & \checkmark & \checkmark & \checkmark & \checkmark & \checkmark & \checkmark & \cite{ALICE:2025cqy} \\
         \cline{3-11}
         & $\phi$ & 0-10\% & $\times$ & \checkmark & \checkmark & \checkmark & \checkmark & \checkmark & \checkmark & \checkmark & \cite{ALICE:2019xyr}\\
         \hline
        \bottomrule
    \end{tabular}

\begin{tablenotes}
\item[a]  Data source contains statistical uncertainties only. Systematic error is assumed to be 5\%, 5\%, 10\% and 15\% of the bin-by-bin value for $\phi$, $\Lambda(\bar{\Lambda})$, $\Xi^{\pm}$ and $\Omega(\bar{\Omega})$ respectively.  
\end{tablenotes}
\end{threeparttable}
\end{table*}

\section{Experimental data}

Particle spectra at mid-rapidity from a wide range of beam energies are readily available. This manuscript closely follows the particle lists collected in Ref.~\cite{Chen:2020zuw}, which focuses on data from STAR~\cite{STAR:2017sal, STAR:2008med, STAR:2019bjj, STAR:2003jwm, Petran:2011aa}, PHENIX~\cite{PHENIX:2013kod} and ALICE~\cite{ALICE:2013mez, ALICE:2013cdo, ALICE:2013xmt, ALICE:2019hno}, but with some minor adjustments. Most prominently, additional $\phi$ and hyperon spectra are now taken from Refs.~\cite{STAR:2015vvs, ALICE:2014jbq, ALICE:2019xyr, ALICE:2025cqy}. The $\phi$ meson, with a mass of \SI{1.019}{GeV/c^2}, is relatively heavy for a meson, which allows us to test whether $q$ correlates more with particle mass or with it being a meson- or baryon. In principle, $J/\psi$ spectra could be useful as the $J/\psi$ particle has a mass of \SI{3.097}{GeV/c^2}, which is heavier than any other hadron considered and could help further explore the behavior of $q$ at extreme masses. However, it is already known that $J/\psi$ does not share the same kinematic values as lighter particles in the context of the Tsallis model~\cite{Shao:2009mu}. Therefore, $J/\psi$ spectra are excluded from this analysis. $\pi^0$ and $K_S^0$ are also not used, except for Au+Au at $\sqrt{s_{NN}}=\SI{62.4}{GeV}$, as they are almost identical in shape to $\pi^{\pm}$ and $K^{\pm}$ and should not affect the freeze-out parameters. For Au+Au at $\sqrt{s_{NN}}=\SI{62.4}{GeV}$, $K_S^0$ is included because $K^\pm$ spectra data are only available at low $p_T$ while the $K_S^0$ spectrum over a wide range of $p_T$ is readily available. The references for all spectra are summarized in~\cref{tab:centDist}.

\Cref{tab:centDist} also lists the centrality-bin assignments used in this analysis. The centrality bin edges are not identical between particles and across beam energies. In general, the finest binning is available for pions, kaons, and protons, whereas the coarsest binning is used for $\phi$ mesons and $\Omega$ baryons. To maintain the most consistent possible particle coverage across centralities while preserving reasonable sensitivity to centrality dependence, this analysis adopts a centrality binning scheme with a granularity that generally lies between the finest and coarsest available choices, taking into account practical considerations such as data availability and species coverage. As a result, pion, kaon, and proton spectra must be merged for some centrality bins. In these cases, statistical uncertainties are propagated in the standard way, while the systematic uncertainty is taken as the average of the merged bins, since it is not reduced by increased statistics. Conversely, the available $\phi$ and hyperon spectra are sometimes reported in bins wider than the fitting bins adopted here. In such cases, those spectra are assigned to the nearest centrality bin and are therefore absent from some bins.

\section{\label{sec:explore}Exploring the relation between non-extensivity parameter and particle species}

To explore the relation between $q$ values and particle species, we fit STAR Au+Au and ALICE Pb+Pb spectra to a custom Tsallis function in which each particle species has an independent $q$ value. Particles and anti-particles are considered identical species in this context and share the same value of $q$. For Au+Au at $\sqrt{s_{NN}}=\SI{62.4}{GeV}$ specifically, $q$ for $K_{S}^{0}$ and $K^\pm$ is considered identical. Due to the significant increase in the number of degrees of freedom (N.D.F.) compared with the previously proposed Tsallis configuration, this new custom Tsallis is not suitable for predicting spectra for unmeasured particles. However, this custom Tsallis is only used for the correlation analysis between $q$ and properties of particle species. When those correlations are found, they can be used to estimate $q$ values for different particle species and allow us to formulate a better fit function.

\begin{figure*}[htp!]
\centering
\includegraphics{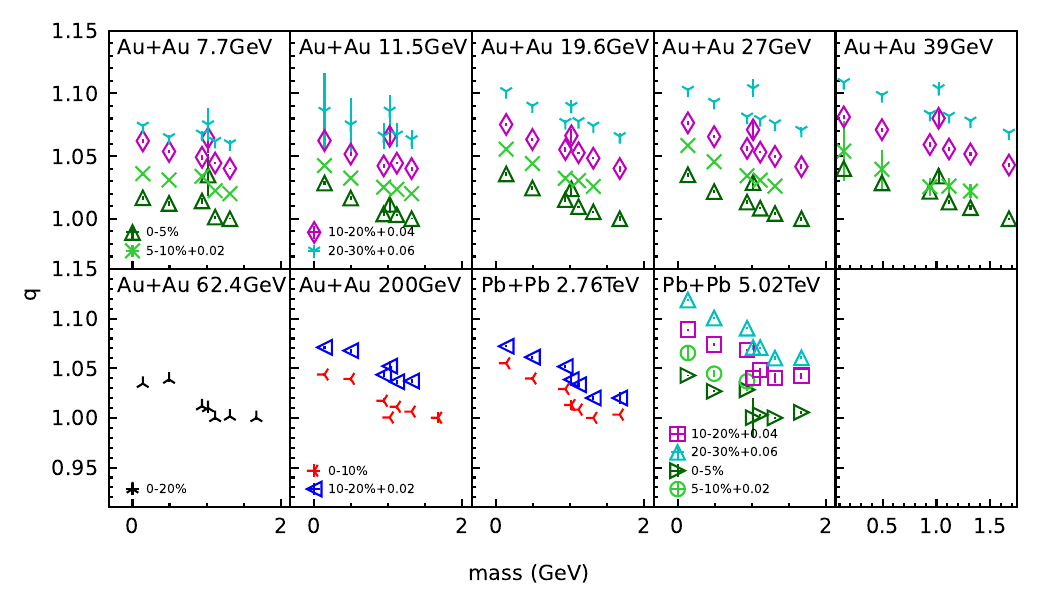}
\caption{$q$ vs. particle masses for the top 30\% centrality bins. $q$ values for different centrality classes are offset by a fixed amount for legibility. Please refer to the legend for the offset values. }
\label{fig:QVsMass1}
\end{figure*}

\begin{figure*}[htp!]
\centering
\includegraphics{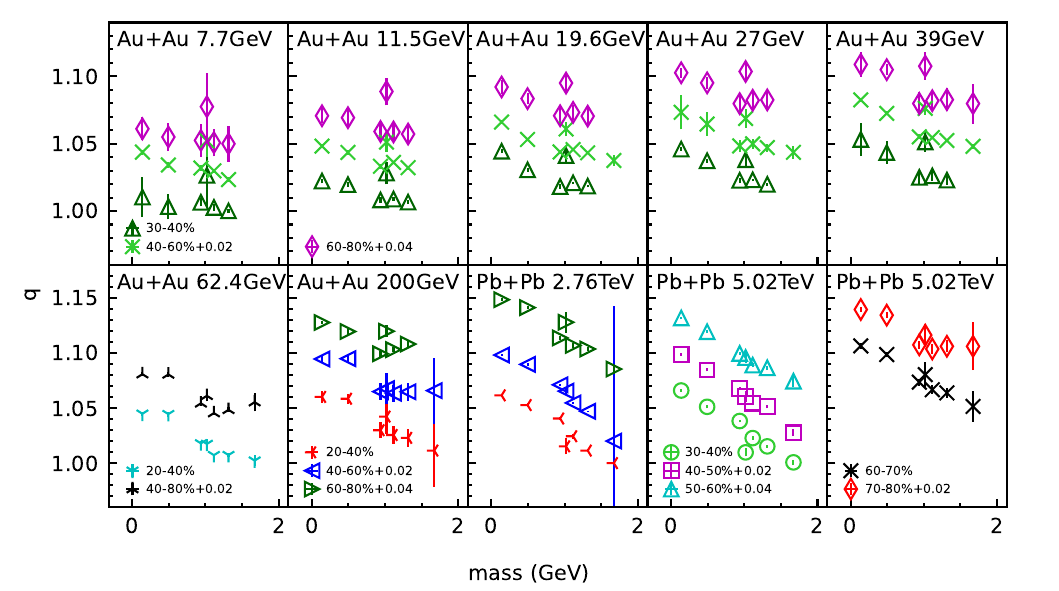}
\caption{Same as~\cref{fig:QVsMass1} for the 30-80\% centrality bins. Note that the two rightmost plots on the lower column both show data for Pb+Pb at $\sqrt{s_{NN}}=\SI{5.02}{TeV}$, but for different centralities. The split is made for legibility. }
\label{fig:QVsMass2}
\end{figure*}

\Cref{fig:QVsMass1} and~\cref{fig:QVsMass2} show the relation between $q$ and the mass of the particle for the top 30\% central and non-central collisions, respectively. It is evident that there is a correlation between $q$ and particle mass, although the correlation strength varies with beam energies and centralities. 

\begin{figure*}[htp!]
\centering
\includegraphics{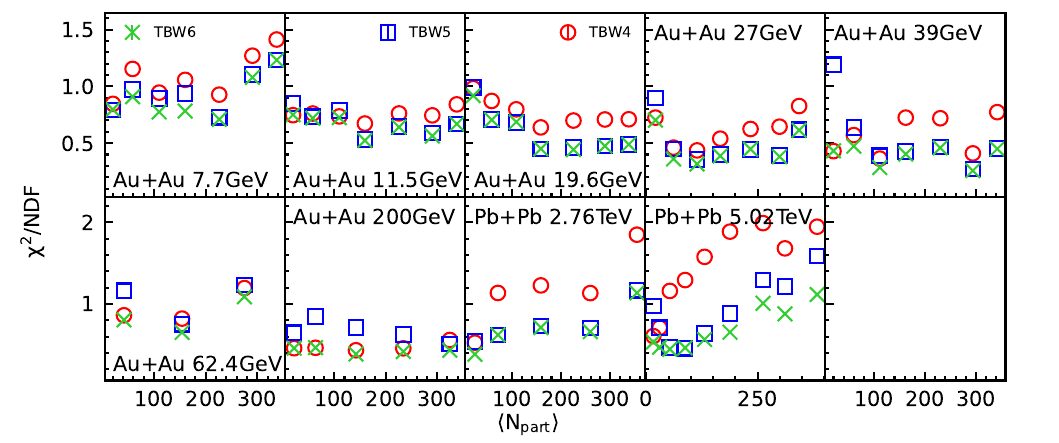}
\caption{$\chi^{2}/{\rm N.D.F.}$ vs. centrality at all beam energies, fitted to all particles in~\cref{tab:particleList}. }
\label{fig:Chi2Ndf}
\end{figure*}

With these results, we propose to parameterize $q$ linearly with the mass of particles. Instead of $q_{\rm meson}$ and $q_{\rm baryon}$ for all mesons and baryons respectively, we propose that $q_i=\max(1, q_{\rm intercept}+m_iq_{\rm slope})$, where $i$ represents the particle species, $m_i$ is the rest mass of particle~$i$, and $q_{\rm intercept}$ and $q_{\rm slope}$ are the two free parameters. We refer to the Tsallis function with this $q$ assignment as TBW5, whereas the previous method of assigning separate $q$ values for mesons and baryons is called TBW4, following the terminology in Ref.~\cite{Chen:2020zuw}.

If we focus on peripheral collisions in~\cref{fig:QVsMass2}, it appears that mesons and baryons do sometimes form two distinct groups for some systems. This is most prominently seen in the 39 GeV data, which clearly shows two groups of $q$ values for peripheral events. If we focus on the 60-80\% result at \SI{39}{GeV} (i.e., the upper rightmost plot of~\cref{fig:QVsMass2}), it clearly shows two distinct groups of $q$ values, with three points on top and three or four on the bottom. The three points with higher $q$ values correspond to pions, kaons, and $\phi$ mesons, and the rest are baryons. This gives credence to the idea of TBW4 and provides early indications that TBW5 will not work perfectly for all systems and centralities. In an attempt to find a fit function that describes as much data as possible, we also propose a more flexible form of the Tsallis function called TBW6, which distributes $q$ values according to

\[
  q= 
\begin{cases}
  \max(1, q^{M}_{\rm intercept}+m_iq_{\rm slope}),& \text{if } i\in (\rm meson)\\
  \max(1, q^{B}_{\rm intercept}+m_iq_{\rm slope}),       & \text{otherwise}.
\end{cases}
\]

where $q^{M}_{\rm intercept}$, $q^{B}_{\rm intercept}$ and $q_{\rm slope}$ are parameters to be fitted. The N.D.F. increases by one compared to TBW4 or TBW5, but this added flexibility is intended to allow TBW6 to describe both central and peripheral spectra at all energies.

\section{\label{sec:results}Results}

\begin{figure*}[htp!]
\centering
\includegraphics{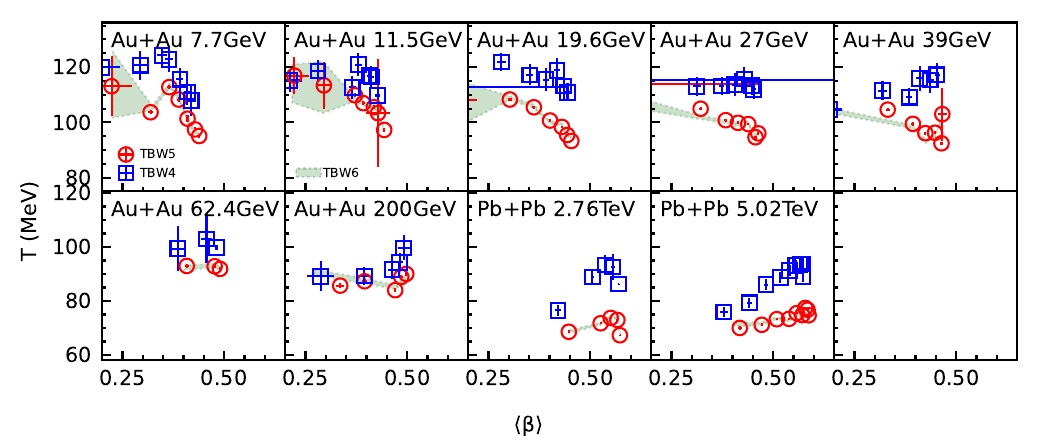}
\caption{$T$ vs. $\langle\beta\rangle$ for all energies fitted to all available particles in~\cref{tab:particleList}. Blue square, red circle markers and green contours correspond to fit to TBW4, TBW5 and TBW6 respectively. }
\label{fig:TsallisParameters}
\end{figure*}

The quality of each fit is measured by $\chi^{2}/{\rm N.D.F.}$ and plotted in \cref{fig:Chi2Ndf} for all centralities and beam energies. In most cases, the three Tsallis variants give $\chi^{2}/{\rm N.D.F.}<1$, especially for data sets between 11.5 and 39 GeV. We believe this is largely driven by an underestimation of $\chi^2$ due to the lack of consideration for bin-by-bin covariance in the systematic uncertainties. Since covariances between $p_T$ points of the spectra are not readily available, this analysis considers uncertainties in each $p_T$ bin as independent, which deflates the $\chi^2$ values. However, it is impossible to rule out the alternative explanation that the Tsallis model overfits data. Consequently, the reduction in $\chi^2$ should be interpreted with caution.

Overall, TBW6 gives the lowest $\chi^{2}/{\rm N.D.F.}$ in most cases. TBW5 performs better than TBW4 in many fits, although TBW5 performs sub-optimally for Au+Au collisions at 62.4 and 200 GeV and in some of the most peripheral centrality bins. A systematic comparison of the three Tsallis variants is given in \cref{tab:statistics2}. In total, TBW5 provides a better fit than TBW4 in 71\% of all cases, while TBW6 outperforms TBW4 in 94\% of them. TBW5 gives better results than TBW4 in more than half of the centrality bins at all energies except 62.4 and 200 GeV. By contrast, TBW6 outperforms TBW4 in more than or equal to 80\% of the centrality bins for every beam energy. If the comparison is restricted to the 30\% most central collisions, both TBW5 and TBW6 almost always achieve a lower \(\chi^{2}/{\rm N.D.F.}\) than TBW4 as~\cref{tab:statistics3} summarizes.

\begin{table}
  \centering
  \caption{Statistics of how TBW5 and TBW6 perform against TBW4. }
  \begin{tabular*}{\linewidth}{@{\extracolsep{\fill}}cccc}
  \toprule
     $\sqrt{s_{NN}}$ (GeV) & N.O. & TBW5 & TBW6 \\
     & Centralities &  is better &   is better\\
     \hline
7.7&7&7(100\%)&7(100\%) \\
11.5&7&5(71\%)&6(86\%) \\
19.6&7&6(86\%)&7(100\%) \\
27&7&6(86\%)&7(100\%) \\
39&7&4(57\%)&6(86\%) \\
62.4&3&1(33\%)&3(100\%) \\
200&5&1(20\%)&4(80\%) \\
2760&5&4(80\%)&5(100\%) \\
5020&9&8(89\%)&9(100\%) \\
\hline
Total&57&42(74\%)&54(95\%) \\
    \bottomrule
  \end{tabular*}
  \label{tab:statistics2}
\end{table}

\begin{table}
  \centering
  \caption{Same as~\cref{tab:statistics2}, but only for the top 30\% centrality bins. }
  \begin{tabular*}{\linewidth}{@{\extracolsep{\fill}}cccc}
  \toprule
     $\sqrt{s_{NN}}$ (GeV) & N.O. & TBW5 & TBW6 \\
     & Centralities &  is better &   is better\\
     \hline
7.7&4&4(100\%)&4(100\%) \\
11.5&4&4(100\%)&4(100\%) \\
19.6&4&4(100\%)&4(100\%) \\
27&4&4(100\%)&4(100\%) \\
39&4&4(100\%)&4(100\%) \\
62.4&1&0(0\%)&1(100\%) \\
200&2&1(50\%)&2(100\%) \\
2760&2&2(100\%)&2(100\%) \\
5020&4&4(100\%)&4(100\%) \\
\hline
Total&29&27(93\%)&29(100\%) \\
    \bottomrule
  \end{tabular*}
  \label{tab:statistics3}
\end{table}

As a visual illustration of the fit results, TBW4, TBW5, and TBW6 fits to central, mid-central, and peripheral data are shown in~\cref{fig:TsallisCentral},~\cref{fig:TsallisMidCentral}, and~\cref{fig:TsallisPeripheral}, respectively. Only Au+Au at 7.7 GeV and Pb+Pb at \SI{5.02}{TeV} are displayed for brevity.

The extracted kinetic-freeze-out temperature \(T\) and average radial flow \(\langle\beta\rangle\) appear in \cref{fig:TsallisParameters}. TBW6 (green band) and TBW5 (red circles) give very similar temperatures, both systematically lower than the values from TBW4 (blue squares). The three models yield comparable flow velocities.

\begin{figure}
\centering
\includegraphics{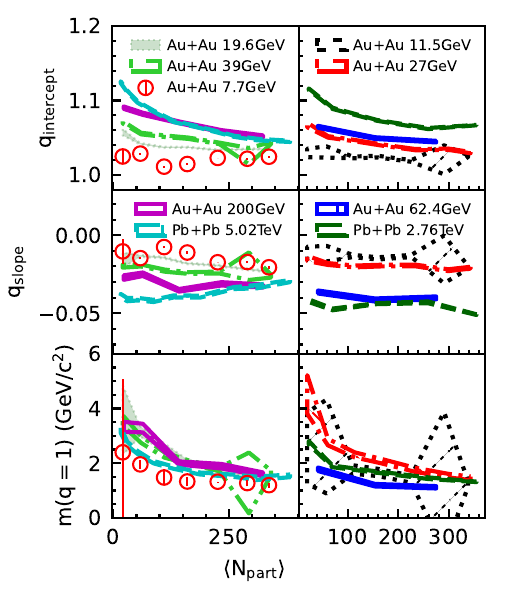}
\caption{Best-fit parameters of the momentum spectrum ($q_{\text{intercept}}$, $q_{\text{slope}}$, and $m(q=1)$) as a function of the average number of participants, $\langle N_{\text{part}} \rangle$, derived from TBW5. The data are split into two columns to improve legibility. Beam energies are distributed across these panels in order, alternating between the left and right sides starting with the lowest energy on the left. Different markers and line styles distinguish between various colliding systems and collision energies, as stated in the legends of the top and middle rows which apply to all subplots.}
\label{fig:qinterceptFunc14}
\end{figure}

From TBW5, we obtain $q_{\rm intercept}$ and $q_{\rm slope}$. Their values are shown in the first two rows of~\cref{fig:qinterceptFunc14}. The slope is always negative, meaning that heavier particles have \(q\) values closer to unity, indicating they appear more equilibrated. This finding contrasts with some earlier expectations~\cite{Bass:2000ib,vanHecke:1998yu,Dumitru:1999sf} and observations~\cite{STAR:2003jis, Barannikova:2004rp, Bass:1999tu, STAR:2005gfr} that hyperons freeze-out earlier than light particles, making them less thermalized during the hadronic stage. The origin of this tension remains unclear. 

One drawback of TBW is that the parameter $T$ loses its intuitive meaning as a temperature. While it represents temperature in ordinary BW, this interpretation fails for non-equilibrium systems where a single temperature does not apply. However, in TBW5, where $q$ depends linearly on mass, we can extrapolate to a particle mass where $q=1$. At this point, TBW reduces to ordinary BW and the intuitive meaning of $T$ is recovered. The extrapolated masses ($m(q=1)$) are shown in the last row of \cref{fig:qinterceptFunc14}. The extrapolated masses indicate that for central collisions with large $\langle N_{\rm part}\rangle$, $m(q=1)\approx\SI{1.4}{GeV/c^2}$ across all beam energies. Thus, the $T$ parameter in TBW5 can be interpreted as the kinetic freeze-out temperature for a hypothetical particle of about \(\SI{1.4}{GeV/c^2}\).

\begin{figure}[!htp]
\centering
\includegraphics{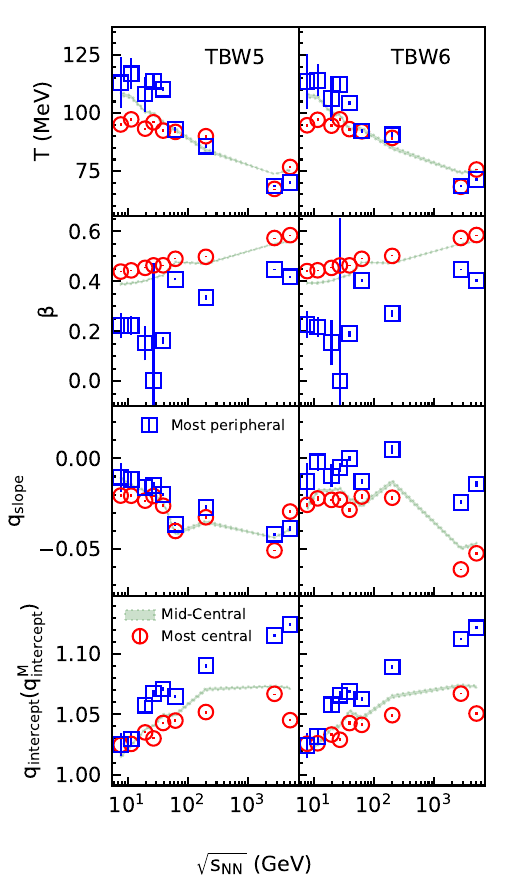}
\caption{Comparison of best-fit parameters as a function of $\sqrt{s_{NN}}$ obtained using TBW5 (left column) and TBW6 (right column). From top to bottom, the panels show $T$, $\beta$, $q_{\text{slope}}$, and the $q$-intercept parameter. Note that for TBW6, the bottom panel displays $q^M_{\text{intercept}}$ specifically, as this function contains multiple $q$-intercept values. To ensure legibility, only data points corresponding to the most central, mid-central, and most peripheral collision classes are shown (legend provided in the third and fourth rows). The specific centrality ranges defining "mid-central" vary by beam energy: 20-30\% for $\sqrt{s_{NN}} = 7.7-\SI{39}{GeV}$; 20-40\% for 62.4-\SI{2760}{GeV}; and 30-40\% for Pb+Pb at \SI{5.02}{TeV}.}
\label{fig:qintercept_energies}
\end{figure}

At the opposite end of the mass dependence, $q_{\rm intercept}$ represents the predicted $q$ value for a hypothetical massless particle. It serves as a metric to quantify and compare overall non-extensivity across beam energies. Plotted against beam energy in the lower left corner of~\cref{fig:qintercept_energies} for most central, mid-central, and most peripheral classes, it shows that non-extensivity increases with beam energy. This is expected since collision duration diminishes at high energies, preventing complete thermalization. The trend is strongest for peripheral collisions, which are less thermalized to begin with due to their smaller fireball size.

The dependence of other TBW5 parameters on beam energy appears in the top three rows of the left column in~\cref{fig:qintercept_energies}. Across all centralities, $T$ decreases steadily with beam energy and increases with impact parameter. $\langle\beta\rangle$ also rises steadily for central to mid-central collisions. The behavior of the most peripheral bins is more erratic, as the fitter sometimes prefers $\langle \beta\rangle=0$, which is the lower bound of that parameter. However, these trends, including that for peripheral events, match observations from previous studies using ordinary BW and TBW~\cite{Chen:2020zuw}. $q_{\rm slope}$ deviates further from zero as beam energy increases, suggesting that different species thermalize at different rates with the disparity enhanced at high energies. It could be related to the fact that different particles take different durations to thermalize, and in high-energy collisions where collision duration is short, lighter particles emerge without being completely thermalized. Of course, this hypothesis assumes that heavier particles thermailze more completely than lighter particles, which, again, contradicts prior observations and expectations~\cite{STAR:2005gfr}. The cause remains unclear.

\begin{figure}
\centering
\includegraphics{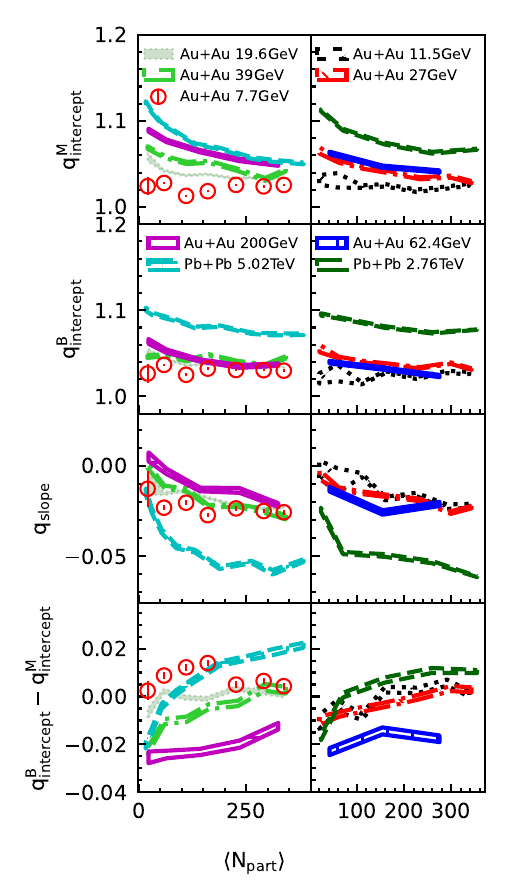}
\caption{Similar to~\cref{fig:qinterceptFunc14}, but showing best-fit parameters derived from the TBW6 fit function instead of TBW5. From top to bottom, the panels display $q^M_{\text{intercept}}$, $q^B_{\text{intercept}}$, $q_{\text{slope}}$, and the difference between the intercepts ($q^B_{\text{intercept}} - q^M_{\text{intercept}}$) as a function of $\langle N_{\text{part}} \rangle$.}
\label{fig:qinterceptFunc15}
\end{figure}

Fitted parameters for TBW6 as a function of $\langle N_{\rm part}\rangle$ appear in \cref{fig:qinterceptFunc15}, and as a function of beam energy in the right column of \cref{fig:qintercept_energies}. Note that the bottom right corner of \cref{fig:qintercept_energies} shows $q_{\rm intercept}^{M}$. The trends in TBW6 closely mirror those in TBW5, such as both $q_{\rm intercept}^{M}$ and $q_{\rm intercept}^{B}$ decreasing with $\langle N_{\rm part}\rangle$ and $q_{\rm slope}$ remaining negative. The bottom row of \cref{fig:qinterceptFunc15} displays the difference between $q_{\rm intercept}^{B}$ and $q_{\rm intercept}^{M}$. In general, this difference becomes more negative as $\langle N_{\rm part}\rangle$ decreases. This indicates that the baryon $q$ values tend to become smaller than the meson values toward more peripheral collisions, suggesting that baryons are usually closer to thermalization than mesons in peripheral collisions, even after the mass dependence is taken into account.

\section{Discussion}

We have investigated two new Tsallis Blast-Wave variants (TBW5 and TBW6) for describing identified‑hadron transverse momentum spectra in relativistic heavy-ion collisions. This analysis provides an early exploration of how $q$ should be ascribed to different particle species, which is an aspect of the Tsallis function that is often overlooked. The mass‑dependent $q$ prescription offers a simple alternative to the binary meson-baryon assignment used in TBW4. TBW5, which assumes a linear dependence of $q$ on particle mass, yields lower $\chi^{2}/{\rm N.D.F.}$ values than TBW4 in 71\% of all datasets across beam energies and centralities, while keeping the number of fit parameters constant. Introducing one additional parameter produces a further refinement, TBW6, which mimics the meson-baryon split by allowing distinct $q$ intercepts for mesons and baryons. This model achieves lower $\chi^{2}/{\rm N.D.F.}$ values than TBW4 in 94\% of all datasets.

Many of our fits produce $\chi^{2}/{\rm N.D.F.} < 1$. This likely reflects the fact that neglecting the covariance between $p_T$ bins in the systematic uncertainties underestimates the true $\chi^{2}/{\rm N.D.F.}$. Consequently, we cannot state with absolute certainty whether the reduction in $\chi^{2}$ values represents genuine improvements or merely overfitting to noise in the $p_T$ spectrum. Given that the improvements occur across most energies and centrality bins, there is reason to believe the improvement is genuine. 

Despite these findings, there is no \textit{a priori} reason for the linear form of the $q$-mass dependence specifically. A more involved microscopic simulation of collision dynamics may reveal a different form of correlation. Nevertheless, linear dependence is the simplest way to parametrize the dependence between variables. The emphasis of this study is to raise the possibility that thermal parameters may differ among particle species in heavy-ion collisions. The detailed mechanism that produces such dependence remains to be studied. It may even be possible that $T$ and $\langle\beta\rangle$ also depend on particle properties in some predictable way. We have refrained from adding too many degrees of freedom in our phenomenological model, but we believe a future study that varies thermal parameters among species would yield further refinements in phenomenological descriptions of particle spectra.

In light of these observations, we propose the following practical guidance:

\begin{enumerate}
    \item For the most central $30\,\%$ of collisions (across all beam energies), TBW5 should be considered a reliable baseline for extracting kinetic freeze‑out parameters, as it typically matches or exceeds the performance of TBW4 while keeping model complexity unchanged.
    
    \item When analyzing peripheral collisions or when the data set includes a very broad range of particle species, TBW6 can be employed to achieve marginally better statistical fits if an increase in degrees of freedom is not an issue; otherwise, TBW4 may be a preferable choice.
\end{enumerate}

\section{Conclusion}

We performed a systematic study of the optimal way to vary the Tsallis non-extensivity parameter $q$ across identified hadrons using STAR Au+Au and ALICE Pb+Pb spectra spanning $\sqrt{s_{NN}} = 7.7-\SI{5.02}{TeV}$. Fitting the data to a custom Tsallis model that uses species-specific $q$ values while sharing a common temperature and velocity profile reveals a clear correlation between $q$ and particle mass, with heavier hadrons approaching $q = 1$. Guided by this empirical trend, we introduced two new Tsallis Blast-Wave models: TBW5, in which $q$ depends linearly on particle mass, and TBW6, which allows the $q$-intercepts for mesons and baryons to differ in addition to the linear dependence on mass. TBW5 outperforms the conventional TBW4 across most centrality classes, while TBW6 provides the most consistent improvements across all systems. These results establish a clear mass ordering in the non-equilibrium behavior at kinetic freeze-out and demonstrate that mass-dependent $q$ assignments yield a more accurate description of identified-hadron spectra from RHIC to LHC energies.

\section{Acknowledgments}
We thank the STAR baryon-junction focus working group for their stimulating discussions leading to this paper. 
This work was supported in part by the Office of Nuclear Physics within the U.S. Department of Energy Office of Science under Contract DE‑FG02‑89ER40531. 

\section{Declaration of generative AI and AI-assisted technologies in the manuscript preparation process}

During the preparation of this work, the authors used Claude Opus for language editing and improving sentence structure. The concepts underlying TBW5 and TBW6 were developed by the authors, who also created the initial analysis code. AI was used only at a later stage to assist with debugging. The authors reviewed and edited all AI-assisted content as necessary and take full responsibility for the content of the published article.

\bibliography{references}

@article{STAR:2017sal,
    author = "Adamczyk, L. and others",
    collaboration = "STAR",
    title = "{Bulk Properties of the Medium Produced in Relativistic Heavy-Ion Collisions from the Beam Energy Scan Program}",
    eprint = "1701.07065",
    archivePrefix = "arXiv",
    primaryClass = "nucl-ex",
    doi = "10.1103/PhysRevC.96.044904",
    journal = "Phys. Rev. C",
    volume = "96",
    number = "4",
    pages = "044904",
    year = "2017"
}

@article{STAR:2008med,
    author = "Abelev, B. I. and others",
    collaboration = "STAR",
    title = "{Systematic Measurements of Identified Particle Spectra in $p p, d^+$ Au and Au+Au Collisions from STAR}",
    eprint = "0808.2041",
    archivePrefix = "arXiv",
    primaryClass = "nucl-ex",
    doi = "10.1103/PhysRevC.79.034909",
    journal = "Phys. Rev. C",
    volume = "79",
    pages = "034909",
    year = "2009"
}

@article{STAR:2003jwm,
    author = "Adams, J. and others",
    collaboration = "STAR",
    title = "{Identified particle distributions in pp and Au+Au collisions at s(NN)**(1/2) = 200 GeV}",
    eprint = "nucl-ex/0310004",
    archivePrefix = "arXiv",
    doi = "10.1103/PhysRevLett.92.112301",
    journal = "Phys. Rev. Lett.",
    volume = "92",
    pages = "112301",
    year = "2004"
}

@article{Tang:2008ud,
    author = "Tang, Zebo and Xu, Yichun and Ruan, Lijuan and van Buren, Gene and Wang, Fuqiang and Xu, Zhangbu",
    title = "{Spectra and radial flow at RHIC with Tsallis statistics in a Blast-Wave description}",
    eprint = "0812.1609",
    archivePrefix = "arXiv",
    primaryClass = "nucl-ex",
    reportNumber = "BNL-KB-02-02",
    doi = "10.1103/PhysRevC.79.051901",
    journal = "Phys. Rev. C",
    volume = "79",
    pages = "051901",
    year = "2009"
}

@article{Shao:2009mu,
    author = "Shao, Ming and Yi, Li and Tang, Zebo and Chen, Hongfang and Li, Cheng and Xu, Zhangbu",
    title = "{Examine the species and beam-energy dependence of particle spectra using Tsallis Statistics}",
    eprint = "0912.0993",
    archivePrefix = "arXiv",
    primaryClass = "nucl-ex",
    doi = "10.1088/0954-3899/37/8/085104",
    journal = "J. Phys. G",
    volume = "37",
    pages = "085104",
    year = "2010"
}

@article{Tang:2011xq,
    author = "Tang, Zebo and Yi, Li and Ruan, Lijuan and Shao, Ming and Chen, Hongfang and Li, Cheng and Mohanty, Bedangadas and Sorensen, Paul and Tang, Aihong and Xu, Zhangbu",
    title = "{Statistical Origin of Constituent-Quark Scaling in the QGP hadronization}",
    eprint = "1101.1912",
    archivePrefix = "arXiv",
    primaryClass = "nucl-ex",
    reportNumber = "BNL-94533-2011-JA",
    doi = "10.1088/0256-307X/30/3/031201",
    journal = "Chin. Phys. Lett.",
    volume = "30",
    pages = "031201",
    year = "2013"
}

@article{Ristea:2013ara,
    author = "Ristea, O. and Jipa, A. and Ristea, C. and Esanu, T. and Calin, M. and Barzu, A. and Scurtu, A. and Abu-Quoad, I.",
    editor = "Li, Bao-An and Natowitz, Joseph",
    title = "{Study of the freeze-out process in heavy ion collisions at relativistic energies}",
    doi = "10.1088/1742-6596/420/1/012041",
    journal = "J. Phys. Conf. Ser.",
    volume = "420",
    pages = "012041",
    year = "2013"
}

@article{Chen:2020zuw,
    author = "Chen, Jia and Deng, Jian and Tang, Zebo and Xu, Zhangbu and Yi, Li",
    title = "{Nonequilibrium kinetic freeze-out properties in relativistic heavy ion collisions from energies employed at the RHIC beam energy scan to those available at the LHC}",
    eprint = "2012.02986",
    archivePrefix = "arXiv",
    primaryClass = "nucl-th",
    doi = "10.1103/PhysRevC.104.034901",
    journal = "Phys. Rev. C",
    volume = "104",
    number = "3",
    pages = "034901",
    year = "2021"
}

@article{Gyulassy:2004zy,
    author = "Gyulassy, Miklos and McLerran, Larry",
    editor = "Rischke, D. and Levin, G.",
    title = "{New forms of QCD matter discovered at RHIC}",
    eprint = "nucl-th/0405013",
    archivePrefix = "arXiv",
    doi = "10.1016/j.nuclphysa.2004.10.034",
    journal = "Nucl. Phys. A",
    volume = "750",
    pages = "30--63",
    year = "2005"
}

@article{STAR:2006egk,
    author = "Adams, J. and others",
    collaboration = "STAR",
    title = "{Scaling Properties of Hyperon Production in Au+Au Collisions at s**(1/2) = 200-GeV}",
    eprint = "nucl-ex/0606014",
    archivePrefix = "arXiv",
    doi = "10.1103/PhysRevLett.98.062301",
    journal = "Phys. Rev. Lett.",
    volume = "98",
    pages = "062301",
    year = "2007"
}

@article{STAR:2015vvs,
    author = "Adamczyk, L. and others",
    collaboration = "STAR",
    title = "{Probing parton dynamics of QCD matter with $\Omega$ and $\phi$ production}",
    eprint = "1506.07605",
    archivePrefix = "arXiv",
    primaryClass = "nucl-ex",
    doi = "10.1103/PhysRevC.93.021903",
    journal = "Phys. Rev. C",
    volume = "93",
    number = "2",
    pages = "021903",
    year = "2016"
}

@article{Becattini:1997rv,
    author = "Becattini, F. and Heinz, Ulrich W.",
    title = "{Thermal hadron production in p p and p anti-p collisions}",
    eprint = "hep-ph/9702274",
    archivePrefix = "arXiv",
    reportNumber = "DFF-268-02-1997",
    doi = "10.1007/s002880050551",
    journal = "Z. Phys. C",
    volume = "76",
    pages = "269--286",
    year = "1997",
    note = "[Erratum: Z.Phys.C 76, 578 (1997)]"
}

@article{schnedermann1994hydrodynamical,
  title={Hydrodynamical assessment of 200A GeV collisions},
  author={Schnedermann, Ekkard and Heinz, Ulrich},
  journal={Physical Review C},
  volume={50},
  number={3},
  pages={1675},
  year={1994},
  publisher={APS}
}

@article{STAR:2019bjj,
    author = "Adam, Jaroslav and others",
    collaboration = "STAR",
    title = "{Strange hadron production in Au+Au collisions at $\sqrt{s_{_{\mathrm{NN}}}}$ = 7.7, 11.5, 19.6, 27, and 39 GeV}",
    eprint = "1906.03732",
    archivePrefix = "arXiv",
    primaryClass = "nucl-ex",
    doi = "10.1103/PhysRevC.102.034909",
    journal = "Phys. Rev. C",
    volume = "102",
    number = "3",
    pages = "034909",
    year = "2020"
}

@article{PHENIX:2013kod,
    author = "Adare, A. and others",
    collaboration = "PHENIX",
    title = "{Spectra and ratios of identified particles in Au+Au and $d$+Au collisions at $\sqrt{s_{NN}}=200$ GeV}",
    eprint = "1304.3410",
    archivePrefix = "arXiv",
    primaryClass = "nucl-ex",
    doi = "10.1103/PhysRevC.88.024906",
    journal = "Phys. Rev. C",
    volume = "88",
    number = "2",
    pages = "024906",
    year = "2013"
}

@article{ALICE:2013mez,
    author = "Abelev, Betty and others",
    collaboration = "ALICE",
    title = "{Centrality dependence of $\pi$, K, p production in Pb-Pb collisions at $\sqrt{s_{NN}}$ = 2.76 TeV}",
    eprint = "1303.0737",
    archivePrefix = "arXiv",
    primaryClass = "hep-ex",
    reportNumber = "CERN-PH-EP-2013-019",
    doi = "10.1103/PhysRevC.88.044910",
    journal = "Phys. Rev. C",
    volume = "88",
    pages = "044910",
    year = "2013"
}

@article{ALICE:2013cdo,
    author = "Abelev, Betty Bezverkhny and others",
    collaboration = "ALICE",
    title = "{$K^0_S$ and $\Lambda$ production in Pb-Pb collisions at $\sqrt{s_{NN}}$ = 2.76 TeV}",
    eprint = "1307.5530",
    archivePrefix = "arXiv",
    primaryClass = "nucl-ex",
    reportNumber = "CERN-PH-EP-2013-132",
    doi = "10.1103/PhysRevLett.111.222301",
    journal = "Phys. Rev. Lett.",
    volume = "111",
    pages = "222301",
    year = "2013"
}

@article{ALICE:2013xmt,
    author = "Abelev, Betty Bezverkhny and others",
    collaboration = "ALICE",
    title = "{Multi-strange baryon production at mid-rapidity in Pb-Pb collisions at $\sqrt{s_{NN}}$ = 2.76 TeV}",
    eprint = "1307.5543",
    archivePrefix = "arXiv",
    primaryClass = "nucl-ex",
    reportNumber = "CERN-PH-EP-2013-134",
    doi = "10.1016/j.physletb.2014.05.052",
    journal = "Phys. Lett. B",
    volume = "728",
    pages = "216--227",
    year = "2014",
    note = "[Erratum: Phys.Lett.B 734, 409--410 (2014)]"
}

@article{ALICE:2019hno,
    author = "Acharya, Shreyasi and others",
    collaboration = "ALICE",
    title = "{Production of charged pions, kaons, and (anti-)protons in Pb-Pb and inelastic $pp$ collisions at $\sqrt {s_{NN}}$ = 5.02 TeV}",
    eprint = "1910.07678",
    archivePrefix = "arXiv",
    primaryClass = "nucl-ex",
    reportNumber = "CERN-EP-2019-208",
    doi = "10.1103/PhysRevC.101.044907",
    journal = "Phys. Rev. C",
    volume = "101",
    number = "4",
    pages = "044907",
    year = "2020"
}

@article{Petran:2011aa,
    author = "Petran, Michal and Letessier, Jean and Petracek, Vojtech and Rafelski, Jan",
    editor = "Bo{\.z}ek, P. and Mr{\'o}wczy{\'n}ski, S",
    title = "{Strangeness Production in Au-Au collisions at $\sqrt{s_{NN}}=62.4$ GeV}",
    eprint = "1112.3189",
    archivePrefix = "arXiv",
    primaryClass = "hep-ph",
    doi = "10.5506/APhysPolBSupp.5.255",
    journal = "Acta Phys. Polon. Supp.",
    volume = "5",
    pages = "255--262",
    year = "2012"
}

@article{De:2007zza,
    author = "De, Bhaskar and Bhattacharyya, S. and Sau, Goutam and Biswas, S. K.",
    title = "{Non-extensive thermodynamics, heavy ion collisions and particle production at RHIC energies}",
    doi = "10.1142/S0218301307006976",
    journal = "Int. J. Mod. Phys. E",
    volume = "16",
    pages = "1687--1700",
    year = "2007"
}

@article{Wilk:2009nn,
    author = "Wilk, Grzegorz and Wlodarczyk, Zbigniew",
    title = "{Multiplicity fluctuations due to the temperature fluctuations in high-energy nuclear collisions}",
    eprint = "0902.3922",
    archivePrefix = "arXiv",
    primaryClass = "hep-ph",
    doi = "10.1103/PhysRevC.79.054903",
    journal = "Phys. Rev. C",
    volume = "79",
    pages = "054903",
    year = "2009"
}

@article{Alberico:1999nh,
    author = "Alberico, W. M. and Lavagno, A. and Quarati, P.",
    title = "{Nonextensive statistics, fluctuations and correlations in high-energy nuclear collisions}",
    eprint = "nucl-th/9902070",
    archivePrefix = "arXiv",
    doi = "10.1007/s100529900220",
    journal = "Eur. Phys. J. C",
    volume = "12",
    pages = "499--506",
    year = "2000"
}

@article{Osada:2008sw,
    author = "Osada, T. and Wilk, G.",
    title = "{Nonextensive hydrodynamics for relativistic heavy-ion collisions}",
    eprint = "0710.1905",
    archivePrefix = "arXiv",
    primaryClass = "nucl-th",
    doi = "10.1103/PhysRevC.77.044903",
    journal = "Phys. Rev. C",
    volume = "77",
    pages = "044903",
    year = "2008",
    note = "[Erratum: Phys.Rev.C 78, 069903 (2008)]"
}

@article{STAR:2004bgh,
    author = "Adams, J. and others",
    collaboration = "STAR",
    title = "{K(892)* resonance production in Au+Au and p+p collisions at s(NN)**(1/2) = 200-GeV at STAR}",
    eprint = "nucl-ex/0412019",
    archivePrefix = "arXiv",
    doi = "10.1103/PhysRevC.71.064902",
    journal = "Phys. Rev. C",
    volume = "71",
    pages = "064902",
    year = "2005"
}

@article{Wilk:1999dr,
    author = "Wilk, G. and Wlodarczyk, Z.",
    title = "{On the interpretation of nonextensive parameter q in Tsallis statistics and Levy distributions}",
    eprint = "hep-ph/9908459",
    archivePrefix = "arXiv",
    reportNumber = "SINS-PVIII-1999-9",
    doi = "10.1103/PhysRevLett.84.2770",
    journal = "Phys. Rev. Lett.",
    volume = "84",
    pages = "2770",
    year = "2000"
}

@article{Wong:2015mba,
    author = "Wong, Cheuk-Yin and Wilk, Grzegorz and Cirto, Leonardo J. L. and Tsallis, Constantino",
    title = "{From QCD-based hard-scattering to nonextensive statistical mechanical descriptions of transverse momentum spectra in high-energy $pp$ and $p\bar p$ collisions}",
    eprint = "1505.02022",
    archivePrefix = "arXiv",
    primaryClass = "hep-ph",
    doi = "10.1103/PhysRevD.91.114027",
    journal = "Phys. Rev. D",
    volume = "91",
    number = "11",
    pages = "114027",
    year = "2015"
}

@article{Urmossy:2011xk,
    author = "Urmossy, Karoly and Barnafoldi, Gergely Gabor and Biro, Tamas Sandor",
    title = "{Generalised Tsallis Statistics in Electron-Positron Collisions}",
    eprint = "1101.3023",
    archivePrefix = "arXiv",
    primaryClass = "hep-ph",
    doi = "10.1016/j.physletb.2011.03.073",
    journal = "Phys. Lett. B",
    volume = "701",
    pages = "111--116",
    year = "2011"
}

@article{ALICE:2019xyr,
    author = "Acharya, Shreyasi and others",
    collaboration = "ALICE",
    title = "{Evidence of rescattering effect in Pb-Pb collisions at the LHC through production of $\rm{K}^{*}(892)^{0}$ and $\phi(1020)$ mesons}",
    eprint = "1910.14419",
    archivePrefix = "arXiv",
    primaryClass = "nucl-ex",
    reportNumber = "CERN-EP-2019-249",
    doi = "10.1016/j.physletb.2020.135225",
    journal = "Phys. Lett. B",
    volume = "802",
    pages = "135225",
    year = "2020"
}

@article{ALICE:2014jbq,
    author = "Abelev, Betty Bezverkhny and others",
    collaboration = "ALICE",
    title = "{$K^*(892)^0$ and $\phi(1020)$ production in Pb-Pb collisions at $\sqrt{s{NN}}$ = 2.76 TeV}",
    eprint = "1404.0495",
    archivePrefix = "arXiv",
    primaryClass = "nucl-ex",
    reportNumber = "CERN-PH-EP-2014-060",
    doi = "10.1103/PhysRevC.91.024609",
    journal = "Phys. Rev. C",
    volume = "91",
    pages = "024609",
    year = "2015"
}

@article{ALICE:2025cqy,
    author = "Abualrob, Ibrahim Jaser and others",
    collaboration = "ALICE",
    title = "{Centrality dependence of strange particle production in Pb-Pb collisions at sNN=5.02 TeV}",
    eprint = "2511.10360",
    archivePrefix = "arXiv",
    primaryClass = "nucl-ex",
    reportNumber = "CERN-EP-2025-256, CERN-EP-2025-254",
    doi = "10.1103/j21y-pqd6",
    journal = "Phys. Rev. C",
    volume = "114",
    number = "2",
    pages = "024909",
    year = "2026"
}

@article{STAR:2007mum,
    author = "Abelev, B. I. and others",
    collaboration = "STAR",
    title = "{Partonic flow and phi-meson production in Au + Au collisions at s(NN)**(1/2) = 200-GeV}",
    eprint = "nucl-ex/0703033",
    archivePrefix = "arXiv",
    doi = "10.1103/PhysRevLett.99.112301",
    journal = "Phys. Rev. Lett.",
    volume = "99",
    pages = "112301",
    year = "2007"
}

@article{STAR:2005gfr,
    author = "Adams, John and others",
    collaboration = "STAR",
    title = "{Experimental and theoretical challenges in the search for the quark gluon plasma: The STAR Collaboration's critical assessment of the evidence from RHIC collisions}",
    eprint = "nucl-ex/0501009",
    archivePrefix = "arXiv",
    doi = "10.1016/j.nuclphysa.2005.03.085",
    journal = "Nucl. Phys. A",
    volume = "757",
    pages = "102--183",
    year = "2005"
}

@article{STAR:2003jis,
    author = "Adams, J. and others",
    collaboration = "STAR",
    title = "{Multistrange baryon production in Au-Au collisions at S(NN)**1/2 = 130 GeV}",
    eprint = "nucl-ex/0307024",
    archivePrefix = "arXiv",
    doi = "10.1103/PhysRevLett.92.182301",
    journal = "Phys. Rev. Lett.",
    volume = "92",
    pages = "182301",
    year = "2004"
}

@inproceedings{Barannikova:2004rp,
    author = "Barannikova, Olga Yu.",
    collaboration = "STAR",
    title = "{Probing collision dynamics at RHIC}",
    booktitle = "{17th International Conference on Ultra Relativistic Nucleus-Nucleus Collisions (Quark Matter 2004)}",
    eprint = "nucl-ex/0403014",
    archivePrefix = "arXiv",
    month = "3",
    year = "2004"
}

@article{Bass:1999tu,
    author = "Bass, S. A. and Dumitru, A. and Bleicher, M. and Bravina, L. and Zabrodin, E. and Stoecker, Horst and Greiner, W.",
    title = "{Hadronic freezeout following a first order hadronization phase transition in ultrarelativistic heavy ion collisions}",
    eprint = "nucl-th/9902062",
    archivePrefix = "arXiv",
    doi = "10.1103/PhysRevC.60.021902",
    journal = "Phys. Rev. C",
    volume = "60",
    pages = "021902",
    year = "1999"
}

@article{Bass:2000ib,
    author = "Bass, S. A. and Dumitru, A.",
    title = "{Dynamics of hot bulk QCD matter: From the quark gluon plasma to hadronic freezeout}",
    eprint = "nucl-th/0001033",
    archivePrefix = "arXiv",
    doi = "10.1103/PhysRevC.61.064909",
    journal = "Phys. Rev. C",
    volume = "61",
    pages = "064909",
    year = "2000"
}

@article{vanHecke:1998yu,
    author = "van Hecke, H. and Sorge, H. and Xu, N.",
    title = "{Evidence of early multistrange hadron freezeout in high-energy nuclear collisions}",
    eprint = "nucl-th/9804035",
    archivePrefix = "arXiv",
    doi = "10.1103/PhysRevLett.81.5764",
    journal = "Phys. Rev. Lett.",
    volume = "81",
    pages = "5764--5767",
    year = "1998"
}

@article{Dumitru:1999sf,
    author = "Dumitru, A. and Bass, S. A. and Bleicher, M. and Stoecker, Horst and Greiner, W.",
    title = "{Direct emission of multiple strange baryons in ultrarelativistic heavy ion collisions from the phase boundary}",
    eprint = "nucl-th/9901046",
    archivePrefix = "arXiv",
    reportNumber = "YRHI-98-26",
    doi = "10.1016/S0370-2693(99)00805-9",
    journal = "Phys. Lett. B",
    volume = "460",
    pages = "411--416",
    year = "1999"
}

@article{STAR:2010yyv,
    author = "Aggarwal, M. M. and others",
    collaboration = "STAR",
    title = "{Strange and Multi-strange Particle Production in Au+Au Collisions at $\sqrt{s_{NN}}$ = 62.4 GeV}",
    eprint = "1010.0142",
    archivePrefix = "arXiv",
    primaryClass = "nucl-ex",
    doi = "10.1103/PhysRevC.83.024901",
    journal = "Phys. Rev. C",
    volume = "83",
    pages = "024901",
    year = "2011",
    note = "[Erratum: Phys.Rev.C 107, 049903 (2023)]"
}

@article{STAR:2008bgi,
    author = "Abelev, B. I. and others",
    collaboration = "STAR",
    title = "{Measurements of phi meson production in relativistic heavy-ion collisions at RHIC}",
    eprint = "0809.4737",
    archivePrefix = "arXiv",
    primaryClass = "nucl-ex",
    doi = "10.1103/PhysRevC.79.064903",
    journal = "Phys. Rev. C",
    volume = "79",
    pages = "064903",
    year = "2009"
}

@article{Heinz:2013th,
  author    = {Heinz, Ulrich W. and Snellings, Raimond},
  title     = {Collective Flow and Viscosity in Relativistic Heavy-Ion Collisions},
  journal   = {Annual Review of Nuclear and Particle Science},
  year      = {2013},
  volume    = {63},
  number    = {1},
  pages     = {123--151},
  doi       = {10.1146/annurev-nucl-102212-170540},
  url       = {https://www.annualreviews.org/content/journals/10.1146/annurev-nucl-102212-170540},
  issn      = {1545-4134},
}

@article{Busza:2018rrf,
  author    = {Busza, Wit and Rajagopal, Krishna and van der Schee, Wilke},
  title     = {{Heavy Ion Collisions: The Big Picture and the Big Questions}},
  journal   = {Annual Review of Nuclear and Particle Science},
  year      = {2018},
  volume    = {68},
  pages     = {339--376},
  doi       = {10.1146/annurev-nucl-101917-020852},
  url       = {https://www.annualreviews.org/content/journals/10.1146/annurev-nucl-101917-020852},
  issn      = {1545-4134},
}

@article{Shuryak:2017rcp,
  author    = {Shuryak, Edward},
  title     = {Strongly coupled quark--gluon plasma in heavy ion collisions},
  journal   = {Reviews of Modern Physics},
  volume    = {89},
  number    = {3},
  pages     = {035001},
  year      = {2017},
  doi       = {10.1103/RevModPhys.89.035001},
  url       = {https://doi.org/10.1103/RevModPhys.89.035001}
}

@article{Lao:2017skd,
    author = "Lao, Hai-Ling and Liu, Fu-Hu and Li, Bao-Chun and Duan, Mai-Ying",
    title = "{Kinetic freeze-out temperatures in central and peripheral collisions: Which one is larger?}",
    eprint = "1703.04944",
    archivePrefix = "arXiv",
    primaryClass = "nucl-th",
    doi = "10.1007/s41365-018-0425-x",
    journal = "Nucl. Sci. Tech.",
    volume = "29",
    number = "6",
    pages = "82",
    year = "2018"
}

@article{Ajaz:2024uvd,
    author = "Ajaz, Muhammad and Shehzad, Majid and Waqas, Muhammad and Alrebdi, Haifa I. and Ahmad, Momhammad Ayaz and Jagnandan, Antalov and Jagnandan, Shawn and Badshah, Murad and Baker, Jalal Hasan and Quraishi, Abdul Mosawir",
    title = "{Multiplicity dependence of the freezeout parameters in high energy hadron-hadron collisions*}",
    eprint = "2402.08535",
    archivePrefix = "arXiv",
    primaryClass = "hep-ph",
    doi = "10.1088/1674-1137/ad2a4c",
    journal = "Chin. Phys. C",
    volume = "48",
    number = "5",
    pages = "053108",
    year = "2024"
}

\clearpage
\onecolumngrid
\appendix
\setcounter{figure}{0}
\setcounter{table}{0}
\renewcommand{\theequation}{S\arabic{equation}}
\renewcommand{\thefigure}{S\arabic{figure}}
\renewcommand{\thetable}{S\arabic{table}}
\section{Tsallis fit to data}

\begin{figure}[htp]
\centering

\subfloat[]{
\includegraphics[]{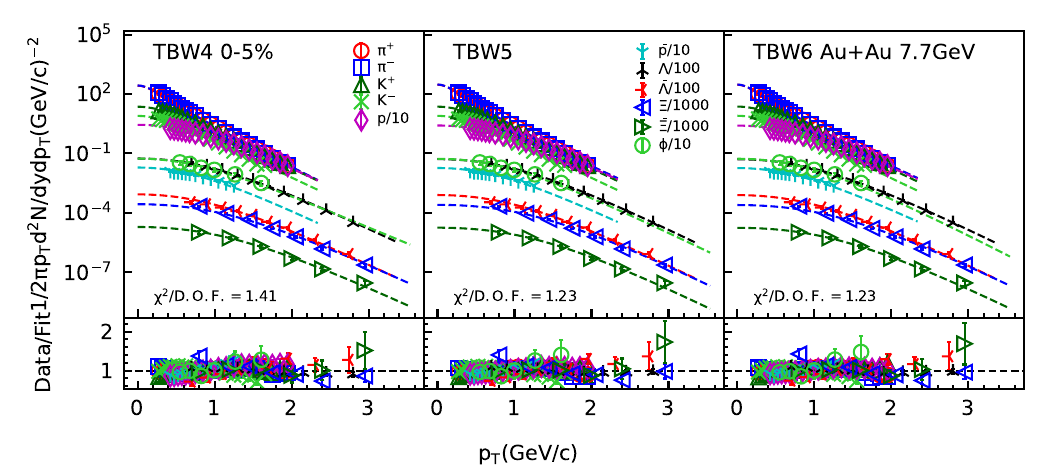}
} \\
\subfloat[]{
\includegraphics[]{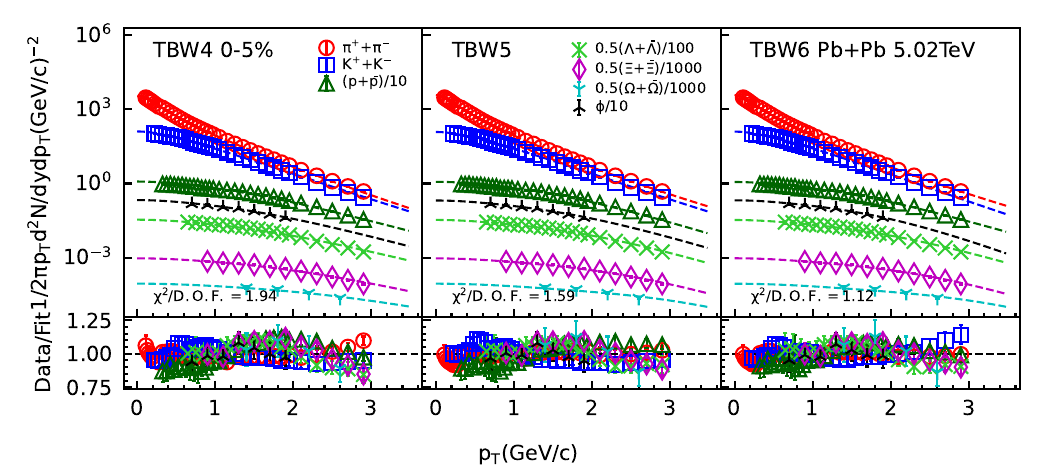}
}
\caption{Fit of TBW4 (left), TBW5 (middle) and TBW6 (right) for the 0-5\% central collisions for (a): Au+Au at $\sqrt{s_{NN}}=\SI{7.7}{GeV}$ and (b): Pb+Pb at $\sqrt{s_{NN}}=\SI{5.02}{TeV}$. The bottom panel in each subfigure shows the ratios of data over predicted values from the best fitted parameters. Note that particle list for (a) and (b) are different. Please refer to the legends for the particle species. }
\label{fig:TsallisCentral}
\end{figure}

\begin{figure}[htp]
\subfloat[]{
\includegraphics[]{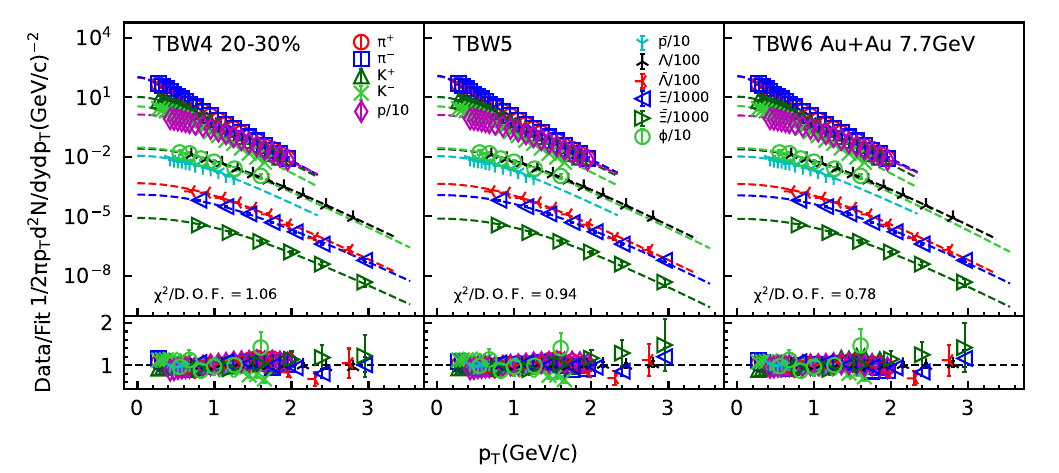}
} \\
\subfloat[]{
\includegraphics[]{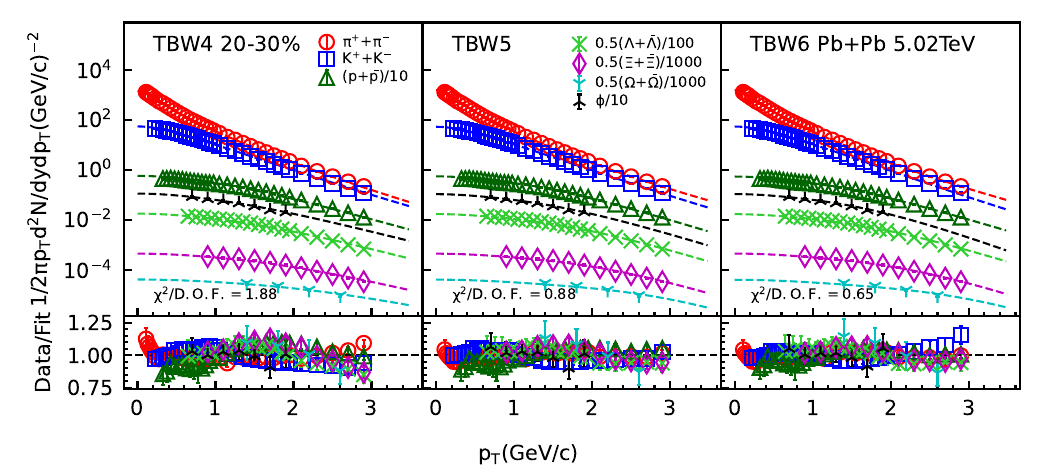}
}
\caption{Same as \cref{fig:TsallisCentral} except for fits to mid-central (20-30\%) collisions.}
\label{fig:TsallisMidCentral}
\end{figure}

\begin{figure}[htp]
\subfloat[]{
\includegraphics[]{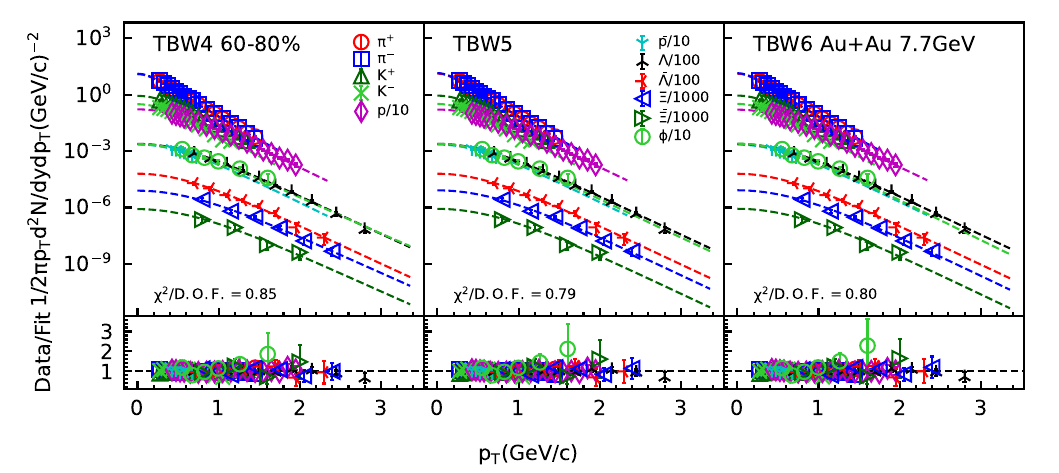}
} \\
\subfloat[]{
\includegraphics[]{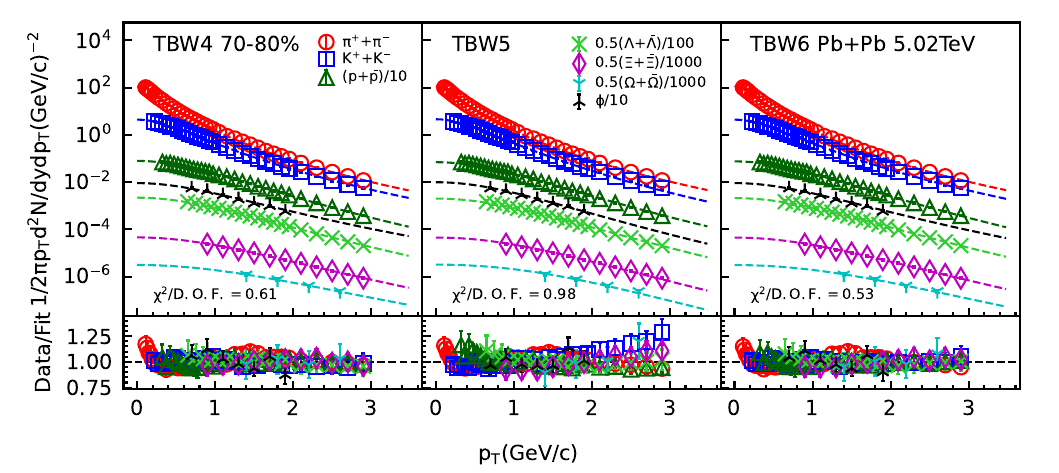}
}
\caption{Same as \cref{fig:TsallisCentral} except for fits to peripheral (60-80\% for Au+Au at 7.7 GeV and 70-80\% for Pb+Pb at 5.02 TeV) collisions.}
\label{fig:TsallisPeripheral}
\end{figure}

\clearpage
\section{Best fitted parameter values}

\begin{table}[htp]
  \centering
  \caption{Best fitted values of kinetic freeze-out parameters and $\chi^2/{\rm N.D.F.}$ from TBW5 when fitted to all data tabulated in \cref{tab:particleList}. }
  \label{tab:TBW4Parameters1}
  \begin{tabular*}{\linewidth}{@{\extracolsep{\fill}}cc|ccccccc}
  \toprule
     System & Centrality & $\langle N_{\rm part}\rangle$ & $\langle \beta\rangle$ & $T$ (MeV) & $q_{\rm intercept}$ & $q_{\rm slope}$ & $m(q=1)$ (GeV/c$^2$) & $\chi^2/\text{DOF}$\\
     \hline 
Au+Au 7.7GeV & 0-5\% & 337.4 & $0.4385(29)$ & $95.1(7)$ & $1.0246(12)$ & $-0.0206(9)$ & $1.19(10)$ & 1.23 \\
 & 5-10\% & 290.4 & $0.4282(32)$ & $97.5(7)$ & $1.0218(13)$ & $-0.0172(9)$ & $1.27(12)$ & 1.10 \\
 & 10-20\% & 226.2 & $0.410(4)$ & $101.3(1.3)$ & $1.022950(6)$ & $-0.017355(5)$ & $1.3223(6)$ & 0.73 \\
 & 20-30\% & 160.2 & $0.388(4)$ & $108.3(8)$ & $1.0147(13)$ & $-0.0111(8)$ & $1.33(18)$ & 0.94 \\
 & 30-40\% & 109.9 & $0.365(5)$ & $112.7(8)$ & $1.0112(14)$ & $-0.0076(8)$ & $1.48(29)$ & 0.89 \\
 & 40-60\% & 58.4 & $0.319(6)$ & $103.7(8)$ & $1.0287(13)$ & $-0.0147(8)$ & $1.95(17)$ & 0.97 \\
 & 60-80\% & 20.2 & $0.22(5)$ & $113(11)$ & $1.025(10)$ & $-0.010(8)$ & $2.4(2.7)$ & 0.79 \\
\hline
Au+Au 11.5GeV & 0-5\% & 338.2 & $0.4434(24)$ & $97.3(6)$ & $1.0255(11)$ & $-0.0206(9)$ & $1.23(9)$ & 0.67 \\
 & 5-10\% & 290.6 & $0.428(29)$ & $103(19)$ & $1.020(19)$ & $-0.016(15)$ & $1.3(2.6)$ & 0.59 \\
 & 10-20\% & 226.0 & $0.4179(30)$ & $105.5(6)$ & $1.0188(11)$ & $-0.0133(7)$ & $1.41(13)$ & 0.65 \\
 & 20-30\% & 159.6 & $0.3919(35)$ & $106.9(7)$ & $1.0247(11)$ & $-0.0149(7)$ & $1.66(13)$ & 0.53 \\
 & 30-40\% & 110.0 & $0.372(4)$ & $109.8(7)$ & $1.0235(12)$ & $-0.0133(7)$ & $1.77(16)$ & 0.79 \\
 & 40-60\% & 58.5 & $0.295(20)$ & $113(9)$ & $1.031(8)$ & $-0.012(5)$ & $2.6(1.7)$ & 0.73 \\
 & 60-80\% & 20.1 & $0.22(4)$ & $117(7)$ & $1.030(6)$ & $-0.012(4)$ & $2.5(1.3)$ & 0.85 \\
\hline
Au+Au 19.6GeV & 0-5\% & 338.0 & $0.4536(20)$ & $93.3(5)$ & $1.0350(8)$ & $-0.0234(6)$ & $1.50(6)$ & 0.49 \\
 & 5-10\% & 289.2 & $0.4434(22)$ & $95.4(5)$ & $1.0346(9)$ & $-0.0218(6)$ & $1.59(7)$ & 0.48 \\
 & 10-20\% & 224.9 & $0.4303(23)$ & $98.4(5)$ & $1.0335(8)$ & $-0.0194(5)$ & $1.73(8)$ & 0.46 \\
 & 20-30\% & 158.1 & $0.4013(25)$ & $100.7(5)$ & $1.0374(8)$ & $-0.0180(5)$ & $2.07(9)$ & 0.45 \\
 & 30-40\% & 108.0 & $0.3615(34)$ & $105.5(6)$ & $1.0372(9)$ & $-0.0143(6)$ & $2.60(15)$ & 0.68 \\
 & 40-60\% & 57.7 & $0.302(5)$ & $108.3(7)$ & $1.0421(10)$ & $-0.0133(6)$ & $3.17(19)$ & 0.71 \\
 & 60-80\% & 19.9 & $0.15(7)$ & $108(8)$ & $1.058(4)$ & $-0.016(5)$ & $3.6(1.3)$ & 0.99 \\
\hline
Au+Au 27GeV & 0-5\% & 343.0 & $0.4634(19)$ & $96.0(5)$ & $1.0300(8)$ & $-0.0208(6)$ & $1.44(7)$ & 0.62 \\
 & 5-10\% & 299.0 & $0.4564(21)$ & $94.6(5)$ & $1.0346(8)$ & $-0.0224(6)$ & $1.54(6)$ & 0.39 \\
 & 10-20\% & 234.0 & $0.4387(21)$ & $99.4(5)$ & $1.0340(8)$ & $-0.0194(5)$ & $1.75(7)$ & 0.44 \\
 & 20-30\% & 166.0 & $0.4134(24)$ & $99.8(5)$ & $1.0404(8)$ & $-0.0195(5)$ & $2.08(8)$ & 0.40 \\
 & 30-40\% & 114.0 & $0.3832(30)$ & $100.8(5)$ & $1.0461(9)$ & $-0.0200(6)$ & $2.30(9)$ & 0.35 \\
 & 40-60\% & 61.2 & $0.322(4)$ & $104.9(6)$ & $1.0512(10)$ & $-0.0185(6)$ & $2.77(12)$ & 0.45 \\
 & 60-80\% & 20.5 & $0.0(5)$ & $113.7(3.0)$ & $1.0672(32)$ & $-0.0149(17)$ & $4.5(7)$ & 0.90 \\
\hline
Au+Au 39GeV & 0-5\% & 342.0 & $0.4637(23)$ & $92.4(6)$ & $1.0428(12)$ & $-0.0262(10)$ & $1.64(9)$ & 0.45 \\
 & 5-10\% & 294.0 & $0.466(11)$ & $103(10)$ & $1.025(11)$ & $-0.020(9)$ & $1.3(1.1)$ & 0.27 \\
 & 10-20\% & 230.0 & $0.4480(24)$ & $96.2(6)$ & $1.0424(10)$ & $-0.0239(6)$ & $1.77(7)$ & 0.47 \\
 & 20-30\% & 162.0 & $0.4241(25)$ & $96.1(5)$ & $1.0494(9)$ & $-0.0242(5)$ & $2.04(7)$ & 0.43 \\
 & 30-40\% & 111.0 & $0.3935(34)$ & $99.5(6)$ & $1.0521(10)$ & $-0.0227(7)$ & $2.30(9)$ & 0.39 \\
 & 40-60\% & 60.0 & $0.332(4)$ & $104.6(7)$ & $1.0556(10)$ & $-0.0194(6)$ & $2.86(12)$ & 0.64 \\
 & 60-80\% & 20.0 & $0.163(15)$ & $110.3(1.0)$ & $1.0709(15)$ & $-0.0199(9)$ & $3.57(19)$ & 1.19 \\
\hline
Au+Au 62.4GeV & 0-20\% & 275.0 & $0.4898(20)$ & $91.9(6)$ & $1.0448(11)$ & $-0.0401(9)$ & $1.12(4)$ & 1.23 \\
 & 20-40\% & 154.2 & $0.4765(23)$ & $92.9(7)$ & $1.0495(11)$ & $-0.0415(10)$ & $1.19(5)$ & 0.75 \\
 & 40-80\% & 41.7 & $0.408(5)$ & $93.0(8)$ & $1.0647(13)$ & $-0.0364(10)$ & $1.78(6)$ & 1.16 \\
\hline
Au+Au 200GeV & 0-10\% & 324.6 & $0.4977(24)$ & $90.0(7)$ & $1.0519(13)$ & $-0.0326(10)$ & $1.59(7)$ & 0.51 \\
 & 10-20\% & 234.3 & $0.4855(30)$ & $88.9(8)$ & $1.0588(14)$ & $-0.0312(11)$ & $1.89(8)$ & 0.63 \\
 & 20-40\% & 142.35 & $0.4705(28)$ & $84.0(7)$ & $1.0707(11)$ & $-0.0353(8)$ & $2.00(6)$ & 0.71 \\
 & 40-60\% & 63.8 & $0.395(6)$ & $87.3(9)$ & $1.0821(14)$ & $-0.0250(11)$ & $3.28(16)$ & 0.85 \\
 & 60-80\% & 22.25 & $0.335(10)$ & $85.7(9)$ & $1.0904(16)$ & $-0.0271(14)$ & $3.33(18)$ & 0.65 \\
\hline
Pb+Pb 2.76TeV & 0-10\% & 354.7 & $0.57296(27)$ & $67.32(12)$ & $1.06698(23)$ & $-0.05090(21)$ & $1.316(8)$ & 1.17 \\
 & 10-20\% & 259.3 & $0.5667(8)$ & $73.0(4)$ & $1.0624(7)$ & $-0.0430(6)$ & $1.449(29)$ & 0.70 \\
 & 20-40\% & 158.3 & $0.5504(5)$ & $73.72(4)$ & $1.0733(5)$ & $-0.04381(31)$ & $1.672(24)$ & 0.73 \\
 & 40-60\% & 70.49 & $0.5257(13)$ & $71.8(4)$ & $1.0894(8)$ & $-0.0475(5)$ & $1.881(30)$ & 0.62 \\
 & 60-80\% & 23.42 & $0.4474(30)$ & $68.6(5)$ & $1.1152(10)$ & $-0.0419(8)$ & $2.75(6)$ & 0.54 \\
\hline
Pb+Pb 5.02TeV & 0-5\% & 383.4 & $0.5844(6)$ & $76.8(4)$ & $1.0451(8)$ & $-0.0294(7)$ & $1.54(5)$ & 1.59 \\
 & 5-10\% & 311.2 & $0.5871(8)$ & $74.7(4)$ & $1.0476(11)$ & $-0.0330(11)$ & $1.45(6)$ & 1.21 \\
 & 10-20\% & 262.0 & $0.5789(7)$ & $77.4(4)$ & $1.0520(8)$ & $-0.0338(6)$ & $1.54(4)$ & 1.29 \\
 & 20-30\% & 187.9 & $0.5714(7)$ & $74.79(35)$ & $1.0638(8)$ & $-0.0394(6)$ & $1.621(35)$ & 0.88 \\
 & 30-40\% & 130.8 & $0.5578(9)$ & $75.5(4)$ & $1.0718(8)$ & $-0.0392(6)$ & $1.83(4)$ & 0.64 \\
 & 40-50\% & 87.14 & $0.5389(12)$ & $73.4(4)$ & $1.0856(8)$ & $-0.0421(7)$ & $2.03(4)$ & 0.46 \\
 & 50-60\% & 52.8 & $0.5088(16)$ & $73.3(4)$ & $1.0977(9)$ & $-0.0410(8)$ & $2.38(5)$ & 0.47 \\
 & 60-70\% & 30.0 & $0.4718(27)$ & $71.2(6)$ & $1.1121(11)$ & $-0.0413(10)$ & $2.72(7)$ & 0.71 \\
 & 70-80\% & 15.8 & $0.418(4)$ & $70.0(6)$ & $1.1246(12)$ & $-0.0388(12)$ & $3.21(10)$ & 0.98 \\
\hline
  \end{tabular*}
\end{table}

\begin{table*}
  \centering
  \caption{Same as \cref{tab:TBW4Parameters1}, but for parameters when data is fitted to TBW4 instead. They should be similar to results from Ref.~\cite{Chen:2020zuw}, but the particle list and centrality bins in this manuscript differ slightly. }
  \label{tab:TBW4Parameters2}
  \begin{tabular*}{\linewidth}{@{\extracolsep{\fill}}cc|cccccc}
  \toprule
  System & Centrality & $\langle N_{\rm part}\rangle$ & $\langle \beta\rangle$ & $T$ (MeV) & $q_{\rm Baryon}$ & $q_{\rm Meson}$ & $\chi^2/\text{DOF}$\\
     \hline 
Au+Au 7.7GeV & 0-5\% & 337.4 & $0.419(11)$ & $108(5)$ & $1.00010(25)$ & $1.0088(34)$ & 1.41 \\
 & 5-10\% & 290.4 & $0.410(9)$ & $111(5)$ & $1.00010(28)$ & $1.0057(28)$ & 1.27 \\
 & 10-20\% & 226.2 & $0.391(8)$ & $116(4)$ & $1.0001(5)$ & $1.0059(23)$ & 0.93 \\
 & 20-30\% & 160.2 & $0.365(11)$ & $123(5)$ & $1.0001(9)$ & $1.000(5)$ & 1.06 \\
 & 30-40\% & 109.9 & $0.347(13)$ & $124.3(3.1)$ & $1.0005(14)$ & $1.0001(20)$ & 0.95 \\
 & 40-60\% & 58.4 & $0.294(16)$ & $121(5)$ & $1.0076(30)$ & $1.010(5)$ & 1.15 \\
 & 60-80\% & 20.2 & $0.20(4)$ & $120(5)$ & $1.014(7)$ & $1.018(8)$ & 0.85 \\
\hline
Au+Au 11.5GeV & 0-5\% & 338.2 & $0.428(6)$ & $109.7(3.2)$ & $1.0001(5)$ & $1.0087(21)$ & 0.84 \\
 & 5-10\% & 290.6 & $0.412(7)$ & $116(4)$ & $1.0001(4)$ & $1.0046(23)$ & 0.74 \\
 & 10-20\% & 226.0 & $0.407(6)$ & $116.9(3.2)$ & $1.000(8)$ & $1.0042(19)$ & 0.76 \\
 & 20-30\% & 159.6 & $0.380(9)$ & $121(4)$ & $1.0026(25)$ & $1.007(4)$ & 0.67 \\
 & 30-40\% & 110.0 & $0.365(9)$ & $113(4)$ & $1.0082(25)$ & $1.019(4)$ & 0.74 \\
 & 40-60\% & 58.5 & $0.281(14)$ & $119(4)$ & $1.0172(24)$ & $1.0250(35)$ & 0.76 \\
 & 60-80\% & 20.1 & $0.211(31)$ & $115(4)$ & $1.019(4)$ & $1.031(5)$ & 0.75 \\
\hline
Au+Au 19.6GeV & 0-5\% & 338.0 & $0.444(5)$ & $110.9(3.2)$ & $1.0001(12)$ & $1.0082(21)$ & 0.71 \\
 & 5-10\% & 289.2 & $0.435(6)$ & $113(5)$ & $1.0011(31)$ & $1.008(6)$ & 0.71 \\
 & 10-20\% & 224.9 & $0.418(7)$ & $119(5)$ & $1.0021(21)$ & $1.006(4)$ & 0.70 \\
 & 20-30\% & 158.1 & $0.391(7)$ & $115(4)$ & $1.0102(23)$ & $1.017(4)$ & 0.64 \\
 & 30-40\% & 108.0 & $0.352(9)$ & $117(4)$ & $1.0162(22)$ & $1.022(4)$ & 0.80 \\
 & 40-60\% & 57.7 & $0.281(12)$ & $121.8(3.1)$ & $1.0234(23)$ & $1.0276(32)$ & 0.87 \\
 & 60-80\% & 19.9 & $0.0(4)$ & $112.8(9)$ & $1.0446(18)$ & $1.0561(18)$ & 0.98 \\
\hline
Au+Au 27GeV & 0-5\% & 343.0 & $0.452(6)$ & $111.9(3.5)$ & $1.0001(6)$ & $1.0068(23)$ & 0.83 \\
 & 5-10\% & 299.0 & $0.447(6)$ & $113.3(3.4)$ & $1.000(4)$ & $1.0066(23)$ & 0.65 \\
 & 10-20\% & 234.0 & $0.429(6)$ & $116(4)$ & $1.0044(26)$ & $1.011(5)$ & 0.62 \\
 & 20-30\% & 166.0 & $0.405(6)$ & $114(4)$ & $1.0117(26)$ & $1.021(5)$ & 0.54 \\
 & 30-40\% & 114.0 & $0.373(7)$ & $113.2(3.4)$ & $1.0180(23)$ & $1.029(4)$ & 0.44 \\
 & 40-60\% & 61.2 & $0.312(10)$ & $113.0(3.5)$ & $1.0269(21)$ & $1.0392(32)$ & 0.46 \\
 & 60-80\% & 20.5 & $0.0(6)$ & $115.3(2.4)$ & $1.0485(13)$ & $1.0615(22)$ & 0.73 \\
\hline
Au+Au 39GeV & 0-5\% & 342.0 & $0.453(6)$ & $117(4)$ & $1.0001(7)$ & $1.0067(31)$ & 0.77 \\
 & 5-10\% & 294.0 & $0.452(4)$ & $117.4(1.1)$ & $1.000(25)$ & $1.0059(22)$ & 0.41 \\
 & 10-20\% & 230.0 & $0.436(8)$ & $115(5)$ & $1.0057(33)$ & $1.015(5)$ & 0.72 \\
 & 20-30\% & 162.0 & $0.410(8)$ & $116(5)$ & $1.0125(30)$ & $1.023(5)$ & 0.72 \\
 & 30-40\% & 111.0 & $0.385(8)$ & $109.1(3.3)$ & $1.0221(26)$ & $1.037(4)$ & 0.36 \\
 & 40-60\% & 60.0 & $0.319(11)$ & $111.5(3.5)$ & $1.0312(25)$ & $1.046(4)$ & 0.57 \\
 & 60-80\% & 20.0 & $0.188(33)$ & $104.7(3.4)$ & $1.046(4)$ & $1.0689(33)$ & 0.43 \\
\hline
Au+Au 62.4GeV & 0-20\% & 275.0 & $0.4811(18)$ & $99.7(9)$ & $1.0001(7)$ & $1.0308(12)$ & 1.19 \\
 & 20-40\% & 154.2 & $0.456(12)$ & $103(9)$ & $1.006(4)$ & $1.036(7)$ & 0.82 \\
 & 40-80\% & 41.7 & $0.386(19)$ & $99(8)$ & $1.0256(32)$ & $1.056(5)$ & 0.86 \\
\hline
Au+Au 200GeV & 0-10\% & 324.6 & $0.491(15)$ & $100(5)$ & $1.012(8)$ & $1.036(5)$ & 0.56 \\
 & 10-20\% & 234.3 & $0.482(9)$ & $94.3(3.3)$ & $1.021(5)$ & $1.048(4)$ & 0.46 \\
 & 20-40\% & 142.35 & $0.462(9)$ & $91.6(3.3)$ & $1.027(5)$ & $1.057(4)$ & 0.43 \\
 & 40-60\% & 63.8 & $0.395(15)$ & $89.2(3.2)$ & $1.051(4)$ & $1.0757(31)$ & 0.46 \\
 & 60-80\% & 22.25 & $0.287(33)$ & $89(6)$ & $1.067(4)$ & $1.0895(30)$ & 0.46 \\
\hline
Pb+Pb 2.76TeV & 0-10\% & 354.7 & $0.56980(27)$ & $86.26(19)$ & $1.00010(20)$ & $1.0273(7)$ & 1.85 \\
 & 10-20\% & 259.3 & $0.556(4)$ & $93(5)$ & $1.016(5)$ & $1.030(6)$ & 1.13 \\
 & 20-40\% & 158.3 & $0.537(4)$ & $93(4)$ & $1.026(4)$ & $1.043(5)$ & 1.23 \\
 & 40-60\% & 70.49 & $0.505(4)$ & $88.9(2.9)$ & $1.042(4)$ & $1.066(4)$ & 1.13 \\
 & 60-80\% & 23.42 & $0.421(7)$ & $76.5(2.1)$ & $1.0754(30)$ & $1.1042(26)$ & 0.53 \\
\hline
Pb+Pb 5.02TeV & 0-5\% & 383.4 & $0.5714(20)$ & $93.6(2.0)$ & $1.0270(35)$ & $1.0236(34)$ & 1.94 \\
 & 5-10\% & 311.2 & $0.5736(21)$ & $89.1(2.0)$ & $1.0292(34)$ & $1.0303(33)$ & 1.68 \\
 & 10-20\% & 262.0 & $0.5663(20)$ & $93.6(2.0)$ & $1.0261(33)$ & $1.0301(33)$ & 1.99 \\
 & 20-30\% & 187.9 & $0.5555(22)$ & $93.1(2.0)$ & $1.0322(30)$ & $1.0399(31)$ & 1.88 \\
 & 30-40\% & 130.8 & $0.5405(24)$ & $91.3(1.9)$ & $1.0399(28)$ & $1.0521(28)$ & 1.58 \\
 & 40-50\% & 87.14 & $0.5176(29)$ & $88.6(1.9)$ & $1.0506(27)$ & $1.0674(27)$ & 1.29 \\
 & 50-60\% & 52.8 & $0.482(4)$ & $86.1(1.9)$ & $1.0644(26)$ & $1.0837(25)$ & 1.16 \\
 & 60-70\% & 30.0 & $0.442(6)$ & $79.3(1.9)$ & $1.0778(29)$ & $1.1031(25)$ & 0.71 \\
 & 70-80\% & 15.8 & $0.379(9)$ & $75.9(1.9)$ & $1.0926(29)$ & $1.1186(22)$ & 0.61 \\
\hline
  \end{tabular*}
\end{table*}

\begin{table*}
  \centering
  \caption{Same as \cref{tab:TBW4Parameters1}, but for parameters when data is fitted to TBW6 instead.}
  \label{tab:TBW6Parameters}
  \begin{tabular*}{\linewidth}{@{\extracolsep{\fill}}cc|ccccccc}
  \toprule
  System & Centrality & $\langle N_{\rm part}\rangle$ & $\langle \beta\rangle$ & $T$ (MeV) & $q^{M}_{\rm intercept}$ & $q^{B}_{\rm intercept}$ & $q_{\rm slope}$ & $\chi^2/\text{DOF}$\\
     \hline 
Au+Au 7.7GeV & 0-5\% & 337.4 & $0.4405(30)$ & $94.7(7)$ & $1.0256(14)$ & $1.0299(14)$ & $-0.0257(13)$ & 1.23 \\
 & 5-10\% & 290.4 & $0.4310(32)$ & $96.6(8)$ & $1.0239(14)$ & $1.0304(15)$ & $-0.0250(13)$ & 1.08 \\
 & 10-20\% & 226.2 & $0.4105(31)$ & $100.2(7)$ & $1.0257(13)$ & $1.0307(13)$ & $-0.0236(12)$ & 0.71 \\
 & 20-30\% & 160.2 & $0.3925(35)$ & $107.5(8)$ & $1.0183(14)$ & $1.0322(14)$ & $-0.0273(13)$ & 0.78 \\
 & 30-40\% & 109.9 & $0.367(4)$ & $113.4(9)$ & $1.0130(15)$ & $1.0251(15)$ & $-0.0204(14)$ & 0.77 \\
 & 40-60\% & 58.4 & $0.323(6)$ & $105.1(9)$ & $1.0279(17)$ & $1.0366(14)$ & $-0.0232(12)$ & 0.91 \\
 & 60-80\% & 20.2 & $0.23(5)$ & $114(12)$ & $1.024(11)$ & $1.027(12)$ & $-0.013(10)$ & 0.80 \\
\hline
Au+Au 11.5GeV & 0-5\% & 338.2 & $0.4437(24)$ & $97.1(7)$ & $1.0260(12)$ & $1.0272(13)$ & $-0.0221(11)$ & 0.67 \\
 & 5-10\% & 290.6 & $0.4295(27)$ & $102.7(7)$ & $1.0226(13)$ & $1.0283(14)$ & $-0.0226(12)$ & 0.56 \\
 & 10-20\% & 226.0 & $0.4183(31)$ & $105.8(7)$ & $1.0189(14)$ & $1.0214(11)$ & $-0.0158(9)$ & 0.64 \\
 & 20-30\% & 159.6 & $0.393(4)$ & $106.9(8)$ & $1.0252(14)$ & $1.0281(12)$ & $-0.0181(10)$ & 0.52 \\
 & 30-40\% & 110.0 & $0.370(4)$ & $108.3(8)$ & $1.0237(14)$ & $1.0152(12)$ & $-0.0051(10)$ & 0.73 \\
 & 40-60\% & 58.5 & $0.294(21)$ & $113(9)$ & $1.031(8)$ & $1.027(10)$ & $-0.008(7)$ & 0.72 \\
 & 60-80\% & 20.1 & $0.22(4)$ & $114(7)$ & $1.032(6)$ & $1.021(6)$ & $-0.002(5)$ & 0.75 \\
\hline
Au+Au 19.6GeV & 0-5\% & 338.0 & $0.4533(21)$ & $94.6(5)$ & $1.0332(11)$ & $1.0337(8)$ & $-0.0229(7)$ & 0.49 \\
 & 5-10\% & 289.2 & $0.4433(24)$ & $95.6(6)$ & $1.0349(12)$ & $1.0368(9)$ & $-0.0237(7)$ & 0.48 \\
 & 10-20\% & 224.9 & $0.4283(25)$ & $99.6(6)$ & $1.0331(12)$ & $1.0364(9)$ & $-0.0218(7)$ & 0.44 \\
 & 20-30\% & 158.1 & $0.4021(26)$ & $99.9(5)$ & $1.0380(11)$ & $1.0371(8)$ & $-0.0176(6)$ & 0.45 \\
 & 30-40\% & 108.0 & $0.361(4)$ & $105.8(6)$ & $1.0366(12)$ & $1.0361(10)$ & $-0.0136(8)$ & 0.69 \\
 & 40-60\% & 57.7 & $0.302(5)$ & $108.8(7)$ & $1.0422(12)$ & $1.0444(11)$ & $-0.0154(9)$ & 0.71 \\
 & 60-80\% & 19.9 & $0.15(9)$ & $106(10)$ & $1.058(5)$ & $1.051(5)$ & $-0.010(6)$ & 0.92 \\
\hline
Au+Au 27GeV & 0-5\% & 343.0 & $0.4629(20)$ & $97.3(5)$ & $1.0289(12)$ & $1.0317(9)$ & $-0.0228(7)$ & 0.61 \\
 & 5-10\% & 299.0 & $0.4566(22)$ & $94.8(6)$ & $1.0352(12)$ & $1.0385(9)$ & $-0.0259(7)$ & 0.38 \\
 & 10-20\% & 234.0 & $0.4383(22)$ & $99.9(5)$ & $1.0333(12)$ & $1.0332(8)$ & $-0.0189(6)$ & 0.45 \\
 & 20-30\% & 166.0 & $0.4140(25)$ & $99.1(5)$ & $1.0406(11)$ & $1.0374(8)$ & $-0.0167(6)$ & 0.39 \\
 & 30-40\% & 114.0 & $0.3840(32)$ & $100.2(6)$ & $1.0454(12)$ & $1.0402(9)$ & $-0.0150(7)$ & 0.32 \\
 & 40-60\% & 61.2 & $0.325(4)$ & $101.7(6)$ & $1.0526(12)$ & $1.0451(10)$ & $-0.0126(8)$ & 0.36 \\
 & 60-80\% & 20.5 & $0.0(7)$ & $112.3(3.5)$ & $1.066(4)$ & $1.056(4)$ & $-0.0049(26)$ & 0.70 \\
\hline
Au+Au 39GeV & 0-5\% & 342.0 & $0.4634(24)$ & $93.1(7)$ & $1.0427(13)$ & $1.0453(14)$ & $-0.0284(12)$ & 0.45 \\
 & 5-10\% & 294.0 & $0.4605(32)$ & $100.6(8)$ & $1.0320(17)$ & $1.0358(15)$ & $-0.0250(12)$ & 0.26 \\
 & 10-20\% & 230.0 & $0.4489(25)$ & $95.2(6)$ & $1.0432(13)$ & $1.0403(10)$ & $-0.0221(7)$ & 0.46 \\
 & 20-30\% & 162.0 & $0.4255(25)$ & $93.1(6)$ & $1.0525(12)$ & $1.0482(9)$ & $-0.0223(6)$ & 0.40 \\
 & 30-40\% & 111.0 & $0.393(4)$ & $98.3(7)$ & $1.0512(13)$ & $1.0416(11)$ & $-0.0134(9)$ & 0.29 \\
 & 40-60\% & 60.0 & $0.336(4)$ & $100.1(7)$ & $1.0582(13)$ & $1.0478(11)$ & $-0.0118(8)$ & 0.47 \\
 & 60-80\% & 20.0 & $0.190(13)$ & $104.3(1.0)$ & $1.0690(15)$ & $1.0463(17)$ & $0.0000(12)$ & 0.43 \\
\hline
Au+Au 62.4GeV & 0-20\% & 275.0 & $0.4898(20)$ & $92.1(7)$ & $1.0413(13)$ & $1.0236(13)$ & $-0.0211(12)$ & 1.09 \\
 & 20-40\% & 154.2 & $0.4756(23)$ & $93.1(7)$ & $1.0468(13)$ & $1.0324(14)$ & $-0.0258(12)$ & 0.65 \\
 & 40-80\% & 41.7 & $0.402(5)$ & $92.3(9)$ & $1.0628(15)$ & $1.0398(15)$ & $-0.0129(13)$ & 0.81 \\
\hline
Au+Au 200GeV & 0-10\% & 324.6 & $0.5013(23)$ & $89.3(8)$ & $1.0493(14)$ & $1.0369(14)$ & $-0.0217(10)$ & 0.43 \\
 & 10-20\% & 234.3 & $0.4908(28)$ & $88.2(8)$ & $1.0547(15)$ & $1.0348(16)$ & $-0.0136(13)$ & 0.42 \\
 & 20-40\% & 142.35 & $0.4725(28)$ & $85.0(7)$ & $1.0648(13)$ & $1.0417(13)$ & $-0.0130(10)$ & 0.39 \\
 & 40-60\% & 63.8 & $0.402(6)$ & $87.3(9)$ & $1.0768(15)$ & $1.0526(19)$ & $-0.0030(15)$ & 0.46 \\
 & 60-80\% & 22.25 & $0.271(14)$ & $90.7(1.1)$ & $1.0892(17)$ & $1.0638(28)$ & $0.0050(25)$ & 0.46 \\
\hline
Pb+Pb 2.76TeV & 0-10\% & 354.7 & $0.57320(27)$ & $68.24(13)$ & $1.0671(6)$ & $1.07757(24)$ & $-0.06141(21)$ & 1.14 \\
 & 10-20\% & 259.3 & $0.5668(9)$ & $73.7(4)$ & $1.0630(11)$ & $1.0739(8)$ & $-0.0533(6)$ & 0.66 \\
 & 20-40\% & 158.3 & $0.5497(11)$ & $74.3(4)$ & $1.0738(11)$ & $1.0806(9)$ & $-0.0496(7)$ & 0.71 \\
 & 40-60\% & 70.49 & $0.5259(13)$ & $71.5(4)$ & $1.0897(10)$ & $1.0903(8)$ & $-0.0483(6)$ & 0.62 \\
 & 60-80\% & 23.42 & $0.4463(30)$ & $68.6(6)$ & $1.1123(11)$ & $1.0952(13)$ & $-0.0242(11)$ & 0.38 \\
\hline
Pb+Pb 5.02TeV & 0-5\% & 383.4 & $0.5849(7)$ & $75.7(4)$ & $1.0507(10)$ & $1.0720(9)$ & $-0.0525(7)$ & 1.12 \\
 & 5-10\% & 311.2 & $0.5899(8)$ & $72.2(4)$ & $1.0537(12)$ & $1.0722(12)$ & $-0.0590(13)$ & 0.88 \\
 & 10-20\% & 262.0 & $0.5799(7)$ & $75.9(4)$ & $1.0570(9)$ & $1.0734(8)$ & $-0.0532(7)$ & 1.01 \\
 & 20-30\% & 187.9 & $0.5731(7)$ & $73.2(4)$ & $1.0680(9)$ & $1.0817(8)$ & $-0.0564(7)$ & 0.65 \\
 & 30-40\% & 130.8 & $0.5587(9)$ & $75.2(4)$ & $1.0730(9)$ & $1.0801(9)$ & $-0.0473(7)$ & 0.57 \\
 & 40-50\% & 87.14 & $0.5397(12)$ & $72.8(4)$ & $1.0866(9)$ & $1.0883(10)$ & $-0.0448(9)$ & 0.46 \\
 & 50-60\% & 52.8 & $0.5079(17)$ & $73.5(5)$ & $1.0969(10)$ & $1.0929(12)$ & $-0.0362(10)$ & 0.45 \\
 & 60-70\% & 30.0 & $0.4666(29)$ & $71.4(6)$ & $1.1105(12)$ & $1.0961(15)$ & $-0.0248(14)$ & 0.47 \\
 & 70-80\% & 15.8 & $0.403(5)$ & $71.4(7)$ & $1.1217(12)$ & $1.1014(18)$ & $-0.0141(17)$ & 0.53 \\
\hline
  \end{tabular*}
\end{table*}

\end{document}